\documentclass[fleqn,usenatbib]{mnras}

\usepackage{newtxtext,newtxmath}

\usepackage{graphicx}
\usepackage{dcolumn}

\usepackage[usenames,dvipsnames]{xcolor}
\hypersetup{colorlinks=true,citecolor=blue,linkcolor=OliveGreen}

\usepackage{soul}

\usepackage{subfigure}
\usepackage{wasysym}
\usepackage{verbatim}
\usepackage{color}
\usepackage{mathtools}
\usepackage{hyperref}
\usepackage{soul}
\usepackage{pbox}
\usepackage[export]{adjustbox}

\usepackage{slashed}
\usepackage{multirow}
\usepackage{booktabs}  
\usepackage{natbib}
\usepackage{float}
\usepackage{afterpage}
\usepackage{dcolumn}
\usepackage{bm}
\usepackage{amssymb,amsmath}  
\usepackage{lipsum}
\usepackage{setspace}
\usepackage{etoolbox}

\newcommand{\Abacus}[1]{\textsc{#1}}

\def\beq{\begin{equation}}
\def\eeq{\end{equation}}

\def\mnras{Mon. Not. R. Astron. Soc}

\usepackage[T1]{fontenc}
\usepackage{ae,aecompl}

\usepackage{graphicx}	
\usepackage{amsmath}	
\usepackage{amssymb}	

\title[Limitations to the bHOD model and beyond]{Limitations to the ``basic'' HOD model and beyond}

\author[B. Hadzhiyska et al.]{
Boryana Hadzhiyska,$^{1}$\thanks{E-mail: boryana.hadzhiyska@cfa.harvard.edu}
Sownak Bose,$^{1}$
Daniel Eisenstein,$^{1}$
\newauthor
\ Lars Hernquist$^{1}$
and David N. Spergel$^{2,3}$
\\
$^{1}$Harvard-Smithsonian Center for Astrophysics, 60 Garden St., Cambridge, MA 02138, USA\\
$^{2}$Department of Astrophysical Sciences, Princeton University, Princeton, NJ 08540 USA\\
$^{3}$Center for Computational Astrophysics, Flatiron Institute, 162 Fifth Avenue, New York, NY 10010, USA\\
}
\date{Accepted XXX. Received YYY; in original form ZZZ}

\pubyear{2019}

\begin{document}
\label{firstpage}
\pagerange{\pageref{firstpage}--\pageref{lastpage}}
\maketitle

\begin{abstract}
We make use of the IllustrisTNG cosmological, hydrodynamical simulations to test fundamental assumptions of the mass-based Halo Occupation Distribution (HOD) approach to modelling the galaxy-halo connection. By comparing the clustering of galaxies measured in the 300 Mpc TNG box (TNG300) with that predicted by the standard (``basic'') HOD model, we find that, on average, the ``basic'' HOD model underpredicts the real-space correlation function in the TNG300 box by $\sim$ 15\% on scales of $1 \ {\rm Mpc}/h < r < 20 \ {\rm Mpc}/h$, which is well beyond the target precision demanded of next-generation galaxy redshift surveys. We perform several tests to establish the robustness of our findings to systematic effects, including the effect of finite box size and the choice of halo finder. In our exploration of ``secondary'' parameters with which to augment the ``basic'' HOD, we find that the local environment of the halo, the velocity dispersion anisotropy, $\beta$, and the product of the half-mass radius and the velocity dispersion, $\sigma^2 R_{\rm halfmass}$, are the three most effective measures of assembly bias that help reconcile the ``basic'' HOD-predicted clustering with that in TNG300. In addition, we test other halo properties such as halo spin, formation epoch and halo concentration. We also find that at fixed halo mass, galaxies in one type of environment cluster differently from galaxies in another. We demonstrate that a more complete model of the galaxy-halo connection can be constructed if we combine both mass and local environment information about the halo.
\end{abstract}




\section{Introduction}

The standard $\Lambda$CDM model predicts that
galaxies form and evolve in virialized structures
made up of dark matter (DM) called halos. The investigation
of
the relationship between galaxies and their parent (host) halos
is of vital importance for constraining fundamental
cosmological parameters and for studying galaxy formation
in detail. This is the case because the dark matter component
is not directly observable, so one way to extract information
about it is by observing the galaxies and their
distributions.
One of the standard and computationally inexpensive approaches 
that is used to study the galaxy distribution is the 
halo occupation distribution
(HOD) model 
\citep{2000MNRAS.318.1144P,2000MNRAS.318..203S,2001ApJ...546...20S,HOD}, 
which determines (probabilistically) the number of
galaxies residing in a host halo 
and assumes that it is governed solely by the halo mass,
remaining agnostic about any other halo property. 
It rests on the long-standing and widely
accepted theoretical prediction that 
halo mass is the attribute that most strongly
influences the halo abundance and halo
clustering as well as the properties of the 
galaxies residing in it \citep{1978MNRAS.183..341W,1984Natur.311..517B}. 
This method provides a framework for ``painting'' a mock galaxy population
on top of large-volume collisionless simulations; i.e. N-body simulations.

The current best approach for probing structure formation on
the largest cosmological scales 
is through N-body simulations. 
They are vital for making
robust predictions
for the upcoming galaxy surveys which will
cover $\sim$ Gpc$^3$ volumes (e.g. Euclid and DESI).
N-body simulations take
substantially less time to evolve compared with hydrodynamical simulations
and can therefore be run on sufficiently large volumes. 
The dark matter halos formed in such a
simulation
are populated with galaxies according to 
different recipes, the most ubiquitously
used one being the standard HOD formalism \citep{2002ApJ...576L.105C,2004MNRAS.350.1153Y,HOD}.
While their properties correlate strongly
with the parent halo mass, galaxies are 
known to be biased tracers of
the halo and total mass
distributions, so other effects need to
be taken into account \citep{Norberg,Zehavi}.
This is known as ``galaxy assembly bias''.
Numerical simulations have shown that 
properties such as
halo formation time, environment, 
concentration, triaxiality, spin, and 
velocity anisotropy play a role in
determining the clustering of halos
\citep{2007MNRAS.377L...5G,2006ApJ...652...71W,
2008PhRvD..77l3514D,2009MNRAS.396.2249W,
2014MNRAS.443.3107L,2006ApJ...652...71W,
2010ApJ...708..469F,2012MNRAS.426L..26L,
2007MNRAS.378..641A,2017A&A...598A.103P,
2018MNRAS.476.5442P,2018MNRAS.473.2486S},
but whether the properties of the galaxies
are correlated with any of these
remains an open question
\citep{2007MNRAS.374.1303C,2019arXiv190811448B}.
Should it turn out to be the case, then the standard
HOD assumption will be violated, and these 
models will likely fail to
predict the clustering statistics 
of galaxies correctly to the necessary
degree of precision. This failure will be particularly
pronounced when trying to create mocks for specially
selected galaxy samples (e.g. on the basis of their color
or star formation rate).

One way to check
whether the modeling for future surveys is done 
at the required levels of precision
is by testing the HOD model against hydrodynamical 
simulations.
Hydrodynamical simulations have now reached a state
where they are sufficiently large in volume and high enough in resolution
that cosmologists can use them to study
the largest structures in our Universe
formed by the elusive dark matter component \citep{2014MNRAS.444.1453D,2015MNRAS.451.1247S,2016MNRAS.463.1797D,2017MNRAS.465.2936M,2018MNRAS.475..676S}.
These simulations can constrain
small-scale physical processes that 
lead to the formation of galaxies
and change their morphology and evolution. 
In addition, they are invaluable for testing various
theoretical models
by comparing their outcome to observations
in the real universe. 

One particular 
set of cosmological simulations
which incorporates state-of-the-art baryonic 
physics models and is useful for probing
the clustering of galaxies
\citep{2005MNRAS.363L..66G,2007MNRAS.377L...5G}
is provided by the IllustrisTNG (TNG) 
team \citep{2018MNRAS.475..676S,2018MNRAS.475..648P,2018MNRAS.475..624N,2018MNRAS.477.1206N,2018MNRAS.480.5113M}.
These models account for a wide range of the
physical processes which are believed to 
govern the formation of galaxies and,
therefore, TNG is well-suited to answer a broad range of questions
regarding how structure in the Universe
evolved over time. In particular, the largest box, TNG300-1
($L_{\rm box} = 205 {\rm \ Mpc}/h$), 
has sufficient volume and resolution to study
clustering of the matter components at relatively
large scales ($\sim$ 20 Mpc$/h$)
(but not sufficiently big to make robust 
predictions for the upcoming galaxy surveys)
and matches well the observed galaxy clustering
\citep{2018MNRAS.475..676S}.

In this paper, we investigate whether there are 
significant violations in the mass-only
HOD model assumptions and what halo properties have the
most significant effect on galaxy clustering according to the
TNG model.
We will refer to this formulation of the HOD as the ``standard''
approach from here on.
The paper is organized as follows: in Section \ref{sec:meth},
we discuss the parameters and specifications of the simulations 
and group finders we have employed as well as the main algorithm
that we follow to test the standard HOD formalism. In Section \ref{sec:res},
we present results from our test of the HOD model in TNG300; we
then check its statistical robustness via N-body only boxes of comparable 
volume, using different
halo definitions. In Section \ref{subsec:sec}, we investigate which
secondary halo properties can
explain the discrepancy we have observed. We then develop a straightforward 
algorithm which allows us to implement an additional partial
dependence of the halo occupation number on an extra halo property.
We finally concentrate on the environmental
dependence and study how the clustering changes when we hold fixed the
type of environment. In Section \ref{sec:conc},
we summarize our results and make elementary proposals for diminishing
the effect of galaxy assembly bias on galaxy assignment recipes in N-body
simulations.

\section{Methods}
\label{sec:meth}
\subsection{Simulations}
In this section, we describe the numerical data used
in this work, providing a brief overview of the 
relevant simulations. Our primary source is the
suite of IllustrisTNG hydrodynamic simulations
and their dark-matter-only counterparts, as summarized in the
TNG data release \citep{2019ComAC...6....2N}\footnote{www.tng-project.org}. We then
test the impact of finite box size and cosmic 
variance in the clustering measured in TNG using 
a much larger N-body simulation volume.

\subsubsection{IllustrisTNG}
The Next Generation Illustris simulation (IllustrisTNG), 
run with the AREPO code \citep{2010MNRAS.401..791S,2019arXiv190904667W}, consists of 9 simulations: 3 box sizes
(300, 100 and 50 Mpc on a side), run at 3 different resolutions 
each \citep{2019MNRAS.tmp.2010N,2019MNRAS.tmp.2024P}. 
IllustrisTNG differs from its predecessor, Illustris 
\citep{2014MNRAS.444.1518V,2014Natur.509..177V,2014MNRAS.445..175G} in that its
sub-grid model has been improved to fix a 
number of shortcomings of the old model: specifically its treatment 
of AGN feedback,  galactic winds and magnetic fields \citep{2018MNRAS.473.4077P,2017MNRAS.465.3291W}.
In addition, there has been some further development of
the numerical implementation concerning the flexibility and
hydrodynamical convergence of the code.

In this work, we use the largest box, 
at its highest resolution, TNG300-1,
a periodic cube of size 205 $\rm{Mpc}/h$ and mass resolution of
$5.9 \times 10^7 M_{\odot}$ and $1.1 \times 10^7 M_{\odot}$
for the dark matter and baryons, respectively.
We take advantage of the fact that
TNG provides
both the hydrodynamical (or full-physics, FP) simulation output 
as well as the dark matter only (or N-body, DMO) one,
 evolved from the same set of initial conditions. This gives us an 
opportunity to make a halo-by-halo assignment of galaxies
by cross-matching the full-physics and dark-matter-only simulations.
The halos (groups) in TNG are found with a standard 
friends-of-friends (FoF) algorithm with linking length $b=0.2$
run on the dark matter particles, while the subhalos are identified 
using the SUBFIND algorithm \citep{Springel:2000qu}, which detects 
substructure within the 
groups and defines locally overdense, self-bound particle groups.

\subsubsection{\Abacus{Abacus}}
We use the publicly released data products produced as part of the AbacusCosmos
N-body simulation suite\footnote{The data products can be found at https://lgarrison.github.io/AbacusCosmos/.}
\citep{Garrison_2018}
to test the statistical robustness of our results, in particular
cosmic variance and box-size effects.
We work with the halo catalog for 
a box of size 720 Mpc$/h$ with 1440$^3$ dark matter particles
which uses $Planck$ 2015 cosmology \citep{2018ApJS..236...43G}.
\Abacus{Abacus} delivers both high speed and accuracy, as it
utilizes novel computational techniques and high 
performance hardware -- e.g. GPUs and RAID disk arrays. 
The force computations are split into a near-field {component},
calculated directly, and a
far-field component, computed 
from  the  multipole
moments  of  particles  in  the  cells. 
The initial conditions are obtained by
scaling back the $z = 0$ power spectrum output from
CAMB \citep{2011ascl.soft02026L} to $z = 49$ and using growth factor
ratios. The initial positions and velocities of the particles
are then generated through
an implementation of 2LPT which includes a rescaling, as the
growing modes near $k_{\rm Nyquist}$ are shown to be suppressed
due to the fact that the dark matter particles are treated 
like macroparticles \citep{2016MNRAS.461.4125G}.
Halos are identified using a nested FoF
halo-finding procedure with two linking length values,
0.186 and 0.117.

\subsection{Procedure}
\label{subsec:algo}
The main assumption of the standard HOD model used to 
populate N-body simulations with galaxies
is that the number of galaxies residing in a 
halo depends
solely on the mass of that halo. Here we test
this conjecture by making use of 
both the IllustrisTNG dark-matter-only output 
as well as the full-physics one. 
Combining these two datasets allows us to draw 
direct statistical
comparisons between the ``truth'' (defined by
the full-physics run) and the HOD model.

{Typically, the HOD model has a functional form with several free parameters which determine the average halo occupation number as a function of mass, $M$.
The model assumes that the number of halos is Poisson sampled from a distribution $N(M)$.}

{
Here, we empirically derive
the galaxy occupation numbers per halo (HOD) by extracting them from 
the TNG full-physics simulation. We then populate the dark-matter-only
halos by randomly shuffling the galaxy occupation numbers in halo mass
bins: this procedure mimics the standard
implementation of the basic HOD because
it preserves mean occupation number as a function of mass, which is
the fundamental assumption of the ``basic'' HOD.}

\begin{figure}
\centering  
\includegraphics[width=0.5\textwidth]{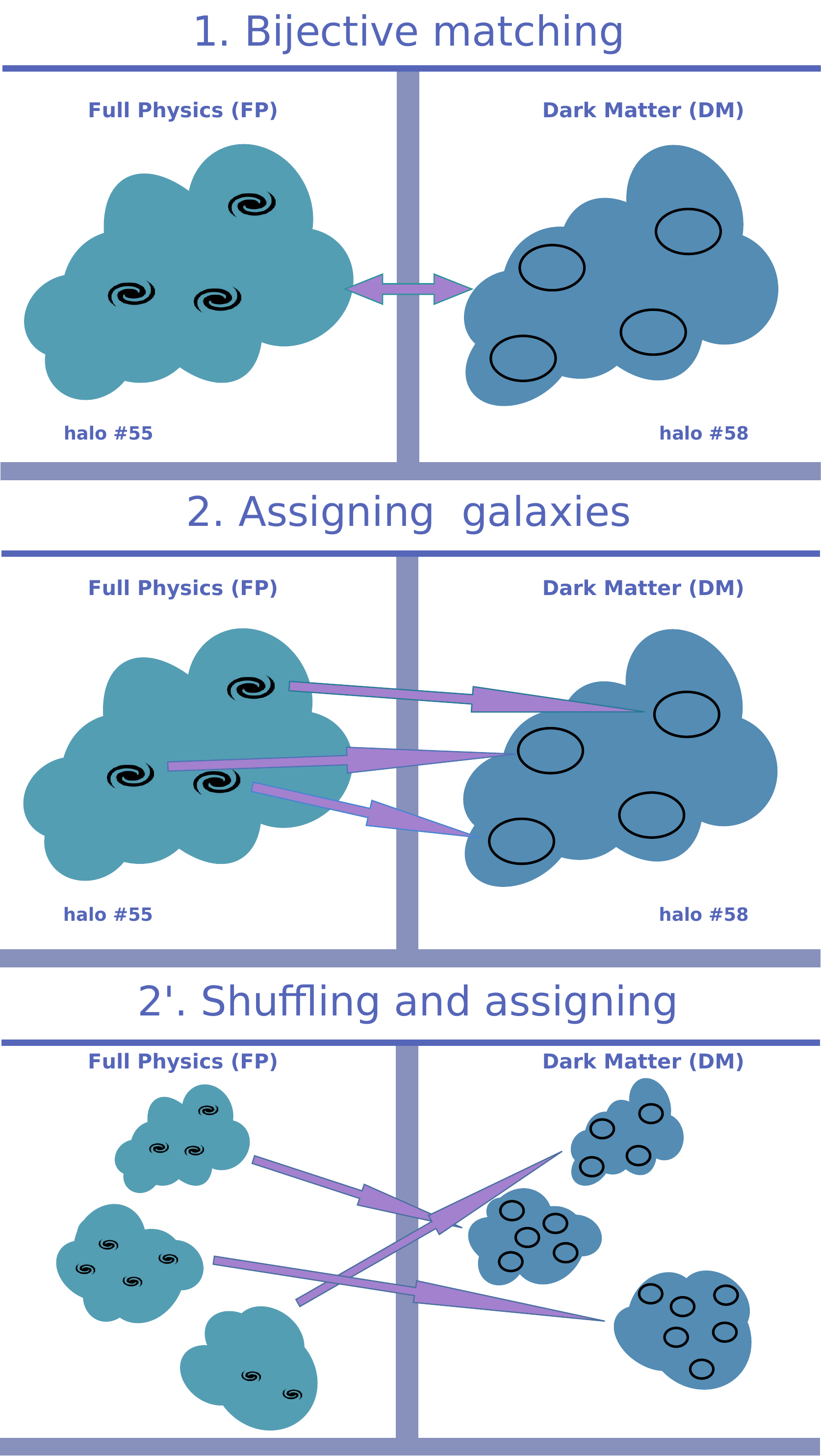}
\caption{Illustration of the procedure outlined in
Section \ref{subsec:algo}. In the first step, we match
the halos in the full-physics (FP) TNG300-1 simulation (\textit{left})
to those in the dark-matter-only (DM) one (\textit{right}) based on
particle IDs. We then have two choices denoted by ``2.''
and  ``2'.'' respectively. We can
either assign the $N$ galaxies that reside in each full-physics halo
to the $N$ subhalos with highest $V_{\rm peak}$ in the
matched dark-matter-only halo or alternatively, 
we can assign those $N$ galaxies to
a different halo belonging to the same mass bin.} 
\label{fig:algo}
\end{figure}

Throughout the paper, we will be referring back to the 
prescription outlined below as the shuffling/ordering
procedure (see Fig. \ref{fig:algo}):
\begin{itemize}
    \item[1.] Bijectively match as many of the halos across the dark-matter-only (DMO) and full-physics (FP) TNG300-1 simulations:\\ A full-physics subhalo is found to be the bijective match of a dark-matter-only subhalo if they share most of each other's particles \citep{2018MNRAS.481.1950L,2019ComAC...6....2N}. If the central
    subhalos of a dark-matter-only and a full-physics halo are bijective matches, then their halo parents are also associated through a bijective match.
    \item[2.] 
    Assign the same number of galaxies to the halos in the dark-matter-only simulation
    as their corresponding full-physics counterparts 
    (as obtained in 1.)
    to obtain the 
    \textbf{fiducial} (\textbf{unshuffled}) galaxy sample. \\ \\
    To create the second, \textbf{shuffled}/\textbf{ordered} 
    sample, instead of assigning 
    the full-physics galaxies to the bijectively
    matched halos, we split the halos into mass bins {such that
    the fractional change within each bin $\frac{M_{\rm max}-M_{\rm min}}{M_{\rm avg}} \lesssim 5\%$.}
    We have checked that the size of these mass bins is
    sufficiently small, so {the HOD shape
    is completely preserved upon shuffling/ordering}. 
    \item[2'.] Order the halos by mass (keeping track of 
    how many galaxies each
    would receive from the bijective match) and, within each mass bin,
    reassign the number of galaxies by either:
    \\
    \textbf{a.} randomly shuffling them \\
    \textbf{b.} ordering them by some halo property (halo concentration, environment, accretion rate).
    
    Note that we
    exclude the most
    massive halos 
    (100 in the case of the TNG300-1 box) because the 
    halo mass function contains very few examples of such high-mass systems.
    \item[3.] The galaxies within a given halo are assigned to the
    subhalos in order of the subhalos' $V_{\rm peak}$ 
    (peak magnitude of the circular velocity attained by the subhalo
    at any point in its evolution); i.e. the subhalos
    with {the} highest $V_{\rm peak}$ gets the first galaxy, 
    the second highest gets the second, etc. There are always
    more subhalos than galaxies that need to be assigned to a given FOF.
    \item[4.] {Split the volume of the simulation into $3^3 = 27$ equal parts and define 27 subsamples} by excluding in each a 
    different cube of side $(205/3) {\rm \ Mpc}/h \approx 68{\rm \ Mpc}/h$
    from the total volume.
    \item[5.] Compute the correlation functions of 
    the shuffled/ordered and fiducial (unshuffled) galaxies in each
    of the 27 
    subsamples using the Landy-Szalay equation \citep{1993ApJ...412...64L}
    \begin{equation}
        \hat \xi_{\rm LS}(r) = \frac{DD(r)}{RR(r)}-1 ,
    \end{equation} 
    assuming periodic boundary conditions.
    \item[6.] To obtain the correlation function and corresponding errors
    for the full box, calculate the mean and jackknife errors of the correlation 
    functions for the 27 subsamples and their ratios {adopting
    the standard equations}
    \begin{equation}
        \overline{\xi(r)}=\frac{1}{n}\sum_{i=1}^{n} \overline{\xi(r)}_i
    \end{equation}
    \begin{equation}
    {\rm Var}(\overline{\xi(r)})=\frac{n-1}{n} \sum_{i=1}^{n} (\overline{\xi(r)}_i - \overline{\xi(r)})^2 ,
    \end{equation}
    {where $n=27$ and $\overline{\xi(r)}_i$ is the correlation
    function value at $r$ for subsample $i$ (i.e. excluding the
    galaxies residing within volume element $i$ in the correlation
    function computation).}
\end{itemize}

\begin{figure}
\centering  
\includegraphics[width=0.5\textwidth]{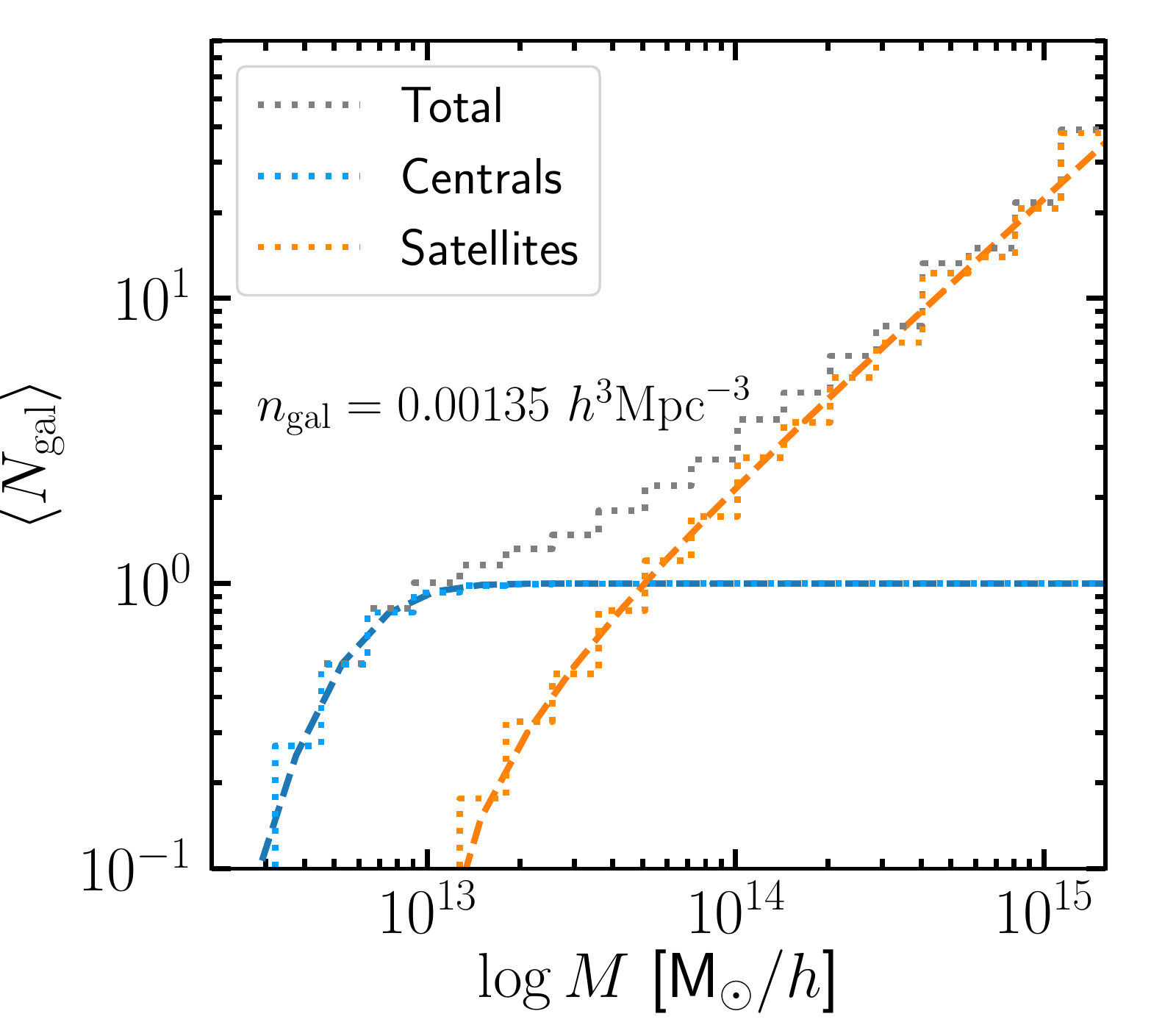}
\caption{Histogram of the average number of galaxies per halo
as a function of halo mass (halo occupation distribution)
in TNG300. Here, we use $M_{\rm 200m}$ as the halo mass 
definition and break the galaxy sample
into two populations -- centrals and satellites.
We also show fits (\textit{dashed line}) to these 
populations assuming the 5-parameter HOD 
model described 
in the {text} \citep{Zheng:2004id}.}
\label{fig:hod_fof}
\end{figure}

\section{Base Model}
\label{sec:res}
\subsection{Shuffled occupations}
To ensure that the galaxy sample from
IllustrisTNG is robust,
we define our galaxies as subhalos with at least
10,000 gravitationally bound
star particles, which results in
a galaxy sample with a number 
density of $n_{\rm gal} \approx 1.3 \times 10^{-3}
\ [{\rm Mpc}/h]^{-3}$.
In Fig. \ref{fig:hod_fof}, we show the HOD 
derived from full-physics TNG300-1 following
the shuffling procedure described in Section 
\ref{subsec:algo}.
  We also fit the 5 
basic HOD parameters from \citet{Zheng:2004id} 
to describe the central and satellite mean occupation functions
\begin{equation}\label{eq:ncen}
    \left< N_{{\rm cen}} (M_{\rm h}) \right> = \frac{1}{2} \left[ 1 + {\rm erf} \left( \frac{\log M_{\rm h}-\log M_{{\rm min}}}{\sigma_{{\log M}}} \right) \right] 
\end{equation}
and
\begin{equation}\label{eq:nsat}
    \left<N_{{\rm sat}} (M_{\rm h})\right> = 
    \left( \frac{M_{\rm h}-M_{{\rm cut}}}{M_1}
    \right)^\alpha ,
\end{equation}
where
$M_{\rm min}$ is the characteristic minimum
mass of halos that host central galaxies,
$\sigma_{\log M}$ is the 
width of this transition, 
$M_{{\rm cut}}$ is the characteristic cut-off
scale for hosting satellites, $M_1$ is a 
normalization factor, and $\alpha$ is the power-law
slope. Our halo mass proxy is $M_{\rm h}=M_{\rm 
200m}$, which is the total mass within a sphere
with mean density 200 times the mean density of the
Universe. Fig. \ref{fig:hod_fof} demonstrates that 
Eqs.~\ref{eq:ncen} and ~\ref{eq:nsat} capture the 
overall shape of the HOD from our simulations very 
well. The corresponding values for the 5 free 
parameters of this model are:
$\log M_{{\rm min}}=12.712$, 
$\sigma_{{\log M}}=0.287$, $\log 
M_{{\rm cut}}=12.95$, $\log M_1 = 13.62$ and 
$\alpha = 0.98$.

Fig. \ref{fig:hod_fof} remains
unchanged after performing
a shuffling of the occupation numbers in 5\% mass
bins following the recipe
outlined in Section \ref{subsec:algo}.
In other words, if the only relevant property
to large-scale galaxy clustering is halo mass,
then the galaxy-galaxy correlation functions in
the shuffled and the unshuffled cases
should be statistically
consistent on large scales, and any deviations
are suggestive of violations of some of our assumptions.

\begin{figure}
\centering 
\includegraphics[width=0.5\textwidth]{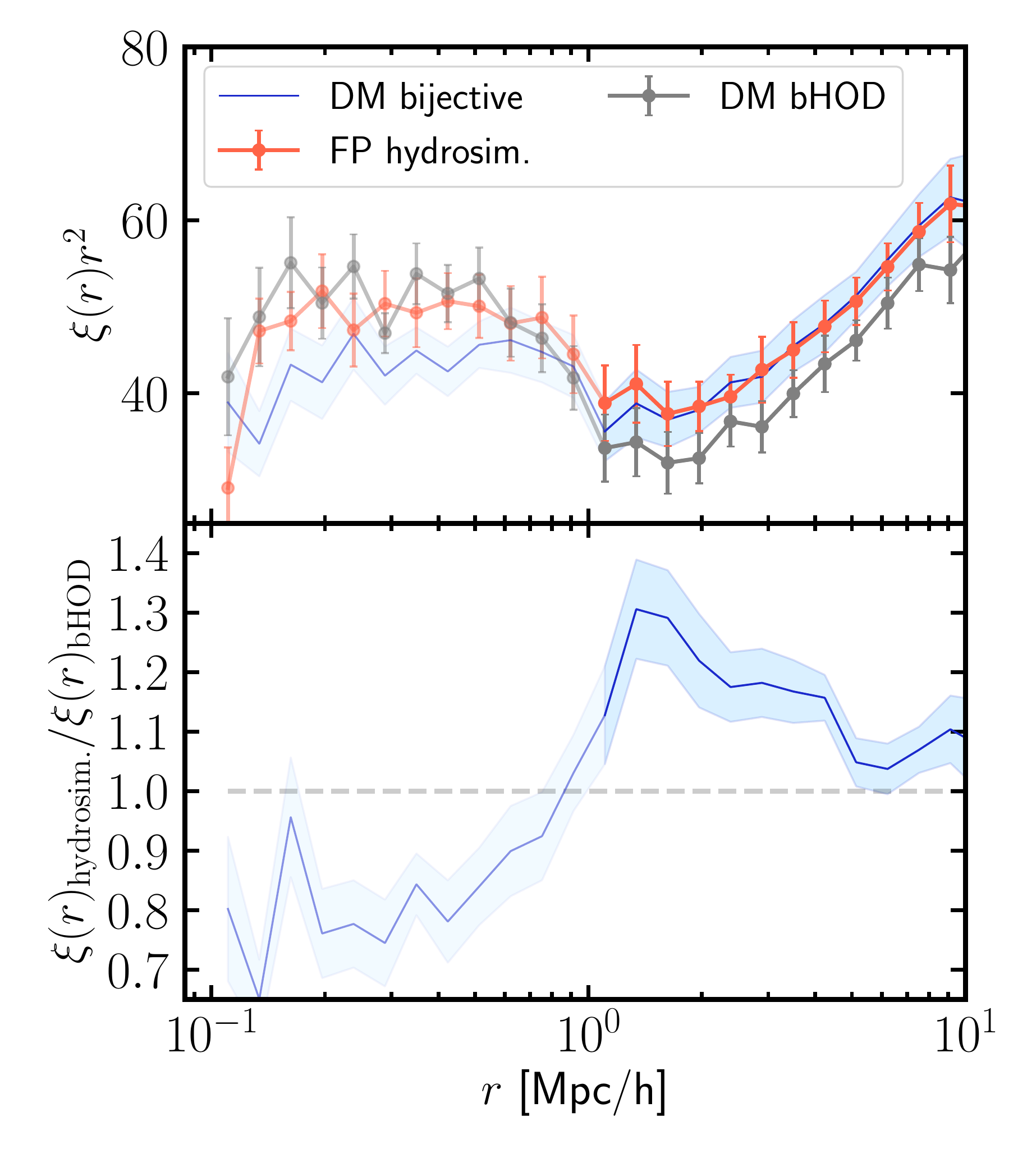}
\caption{
Correlation functions of the full-physics galaxies (FP hydrosim.), the bijective matches (DM bijective), 
and those shuffled in 5\% bins (DM bHOD) (\textit{upper panel}). The ratio
of bijectively matched to shuffled (\textit{lower panel}) shows 
a big discrepancy on large scales of 15\%, which should not 
exist if mass is the only halo property that determines the 
average halo occupation. All errors are jackknife.
This 15\% deviation persists even if we relax the definition of
a galaxy from a subhalo containing at least 
10000 bound star particles to one containing at least 5000.}
\label{fig:shuff_rat}
\end{figure}

The top panel of Fig. \ref{fig:shuff_rat} shows the 
correlation function
of the ``true'' galaxy distribution in the full-physics simulation run of
TNG300-1 in orange, the bijectively matched ones
in blue and the shuffled case, i.e. mimicking
the HOD model, in gray. 
The bottom plot shows
the ratio between the blue and gray curves.
The proxy used for halo mass
here is $M_{\rm 200m}$. 

Fig. \ref{fig:shuff_rat} shows that the clustering of
the full-physics galaxies on scales
above 1 Mpc/h is substantially larger, 10-30\%, than those of the
shuffled dark matter.  This is the key result of our paper, as it
indicates a clear violation of the assumptions of the basic HOD model.
We evaluate this discrepancy by  averaging
the percentage difference for $20 \ {\rm Mpc}/h > r > 1 \ {\rm Mpc}/h$
and find it to be $15 \pm 1\%$ (accounting for the random variation
when shuffling).
In the subsequent sections,
we will attempt to determine the cause of this difference
as well as subject it to more rigorous testing.
The difference
between bijectively matched curve (blue) and 
the full-physics curve (orange) on scales below $r \leq 1 \ {\rm Mpc}/h$
is purely a consequence of how we choose to populate
the central and satellite galaxies within
the halo which is not a subject of study in this
paper. We therefore attribute
no particular importance to the 
differences on such scales.

\subsection{Choice of halo finder}
An often overlooked potential source of discrepancy in N-body simulations, 
apart from the lack of baryonic physics, is the
implementation of halo-finding algorithms, which aim
to identify bound structures (halos) using information about 
the distribution of particles in the simulation over time or
at a particular snapshot.
A very commonly used and relatively fast approach employs
the friends-of-friends (FoF) algorithm, which links together
all particles whose mutual separation is less than the
so called linking length parameter, $b$. 
A problem with the FoF alogorithm is `percolation', in
which many fragmented objects can be linked together as one, 
resulting in unphysical objects entering the halo catalog. 
This pathology could have an impact on statistical 
properties like halo abundance and clustering. Furthermore, 
it can bias weak-lensing estimates of 
cluster masses at a level comparable to the precision of the
most advanced experiments 
and lead to differences in e.g. the halo mass function
beyond the few percent precision necessary for 
cluster  abundance  experiments such as SDSS, DES, DESI, Euclid, and LSST
\citep{2019arXiv190301709G}. 
We are interested in
the effect of the FoF algorithm on the galaxy clustering
in TNG, as the
the choice of linking length can result in different 
structures being linked together (or not), and might 
therefore affect the dark-matter-only and full-physics halo catalogs 
differently due to chance fragmentations.

Algorithms that identify halos {or analyze the particle trajectories
using full 6D phase-space information (i.e. positions and
velocities) such as ROCKSTAR and SPARTA may be better at circumventing these halo pathologies.
They are typically run in post-processing
and can provide alternative halo catalogs to standard
on-the-fly implementations.}
One can conjecture that 
the observed discrepancy on large scales
in Fig. \ref{fig:shuff_rat} is due to the FoF
catalog, 
containing an excess of smaller halos on the 
outskirts of larger groups if the linking length
is too small. Then, as we reshuffle, we place
galaxies in halos which should have been part
of other larger halos, whose mass has been
undermined due to this effect of overshredding.
This results in a suppression of the correlation 
function on large scales.

\begin{figure}
\centering  
\includegraphics[width=0.5\textwidth]{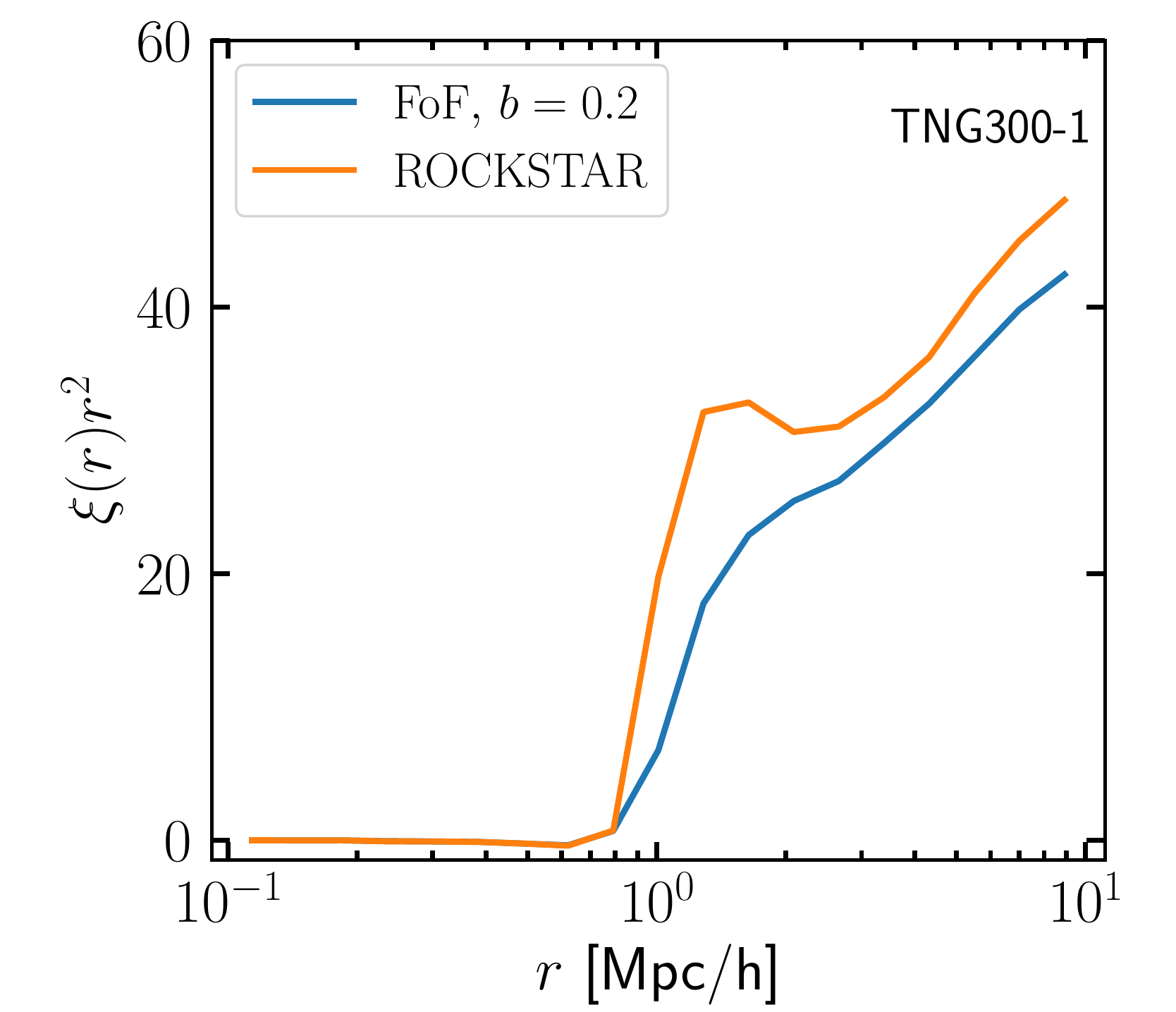}
\caption{Cross correlation between
the 400 most massive and 
the 500,000 most massive
halos in TNG300-1-Dark as defined by ROCKSTAR
and FoF with linking length $b = 0.2$. 
For a mass proxy we use $M_{\rm 200m}$.
The bump around $r = 1{\rm Mpc}/h$ shows
that there are more intermediate-size
halos found near the boundaries 
of the largest halos in the ROCKSTAR
case compared with the FoF.}
\label{fig:rock_fof}
\end{figure}

To test this conjecture, 
we run ROCKSTAR, a phase-space, temporal halo 
finder \citep{2013ApJ...762..109B},
on the final state ($z = 0$) of the TNG300-1-Dark
box. In Fig. \ref{fig:rock_fof}, we show the 
cross-correlation between the 400 most massive and 
the 500,000 most massive parent halos 
in both the {(spatial-only)} FoF and the ROCKSTAR
halo catalogs {(effectively a phase-space FoF finder)},
using $M_{\rm 200m}$ and the virial mass $M_{\rm vir}$,
respectively, as defined by each 
catalog. Contrary to our expectation,
we find an excess of halos separated by 
$\sim 1$ Mpc$/h$ in the ROCKSTAR catalog
relative to the FoF one {(which we also observed
for different mass-scale choices of the two sets of halos being cross-correlated)}. 
This suggests that there is \textit{less}
overshredding in the FoF halo sample
with ROCKSTAR tending to find more 
satellite halos orbiting at the
outskirts of larger halos than
FoF. Indeed, as Table 
\ref{table:proxy} suggests,
the use of the ROCKSTAR catalog
does not alleviate
the discrepancy illustrated in Fig. \ref{fig:shuff_rat}, and we
still find a difference on large
scales of order 15\%.

Another possible source of error is
the definition of halo mass. It has 
been proposed 
\citep{2014ApJ...789....1D,2014JCAP...11..019A,2015ApJ...810...36M,sparta}
that a more
physical definition of a halo boundary
is the ``splashback radius'', defined as 
the apocenter of all particles that ever
fall into the potential well of what is 
ultimately defined as the halo.
We have compared the TNG FoF catalog
with one augmented with {\sc Sparta},
an algorithm designed for computing
splashback radii, and have
confirmed that the TNG FoF
catalog
provides a value for $M_{\rm 200m}$ comparable
to the splashback mass, $M_{\rm sp} \lesssim 2 M_{\rm 200m}$,
obtained by \citet{sparta},
in the mass range of our interest at $z = 0.1$ (see also
Fig. 3 in \citet{2017ApJ...843..140D} {which shows a similarity
between the masses and radii at $z = 0$).
The scatter in the relation between 
$M_{\rm sp}$ and $M_{\rm 200m}$
depends prominently on the accretion rate of halos which
is directly related to the halo clustering. Despite
the similarity between $M_{\rm 200m}$ and $M_{\rm sp}$,
we would need to test more robustly the conjecture 
that using either as a mass proxy in the HOD model
would lead to equivalent results in the case of TNG300}
\citep{2019arXiv190200030M}.

We explore other 
popular mass definitions such as
$M_{\rm 200c}$ and $V_{\rm peak}$.
The procedure for reshuffling is the
same as outlined in Section 
\ref{subsec:algo}, the only difference
being the mass proxy used for creating 
the 5\% mass bins. The results, shown
in Table \ref{table:proxy}, indicate
that indeed defining halo mass as
$M_{\rm 200m}$ leads to the least amount of discrepancy on large scales
($r = 1-20$ Mpc$/h$), of 
$12 \pm 1$\%, out of the commonly used mass proxies.

Out of all definitions, the discrepancy is 
{smallest, $8 \pm 1\%$},
when we adopt $M_{\rm FoF}$. 
We believe that this is most likely due to the
play of several different FoF properties. In particular, we
point out the tendency of the
FoF algorithm to join together
several subhalos which may not
be gravitationally bound to the
cluster but are located 
in close proximity to it.
In this way, the correlation
between halo mass 
$M_{\rm FoF}$ and number of
satellites is
strengthened and our reordering
using $M_{\rm FoF}$ becomes
effectively a reordering on
the number of satellites, which 
by construction increases the 
clustering of satellite
galaxies \citep{2014MNRAS.442.1930P}. This result can be
interpreted either as an
indication that the halo structure
extends beyond the conventional
spherical radius of
$R_{\rm 200 m}$ for some objects
or as an indication of possible
anomalies with the FoF algorithm.

\begin{table*}
\centering
\begin{tabular}{||c | c | c| c||} 
 \hline
 {Mass proxy} & Mass proxy definition used in bHOD & Compared against & Difference from bHOD  \\ [0.5ex] 
 \hline\hline
 $M_{\rm 200m}$  & \pbox{20cm}{\vspace{0.1cm}The total mass enclosed in a sphere with mean
density \\ 200 times the mean density of the 
Universe \vspace{0.1cm}} & hydrosimulation & $15\pm 1 \%$ \\
 \hline
 $M_{\rm vir}$  & Total particle mass (within the virial radius) in ROCKSTAR & hydrosimulation & $15\pm 1$\% \\
 \hline
 $M_{\rm 200c}$ & \pbox{20cm}{\vspace{0.1cm}Total Mass enclosed in a sphere 
 with mean density \\ 200 times the critical density of 
 the Universe \vspace{0.1cm}} & hydrosimulation & $19\pm 1 \%$ \\
 \hline
 $M_{\rm 500c}$ & \pbox{20cm}{\vspace{0.1cm}Total Mass enclosed in a sphere
 with mean density \\ 500 times the critical density of the
 Universe  \vspace{0.1cm}} & hydrosimulation & $19\pm 1 \%$ \\
 \hline
  $\sigma^2 R_{\rm halfmass}$ & 
 \pbox{20cm}{\vspace{0.1cm}Dispersion velocity times the radius containing
 \\ half of the total mass of the largest subhalo \vspace{0.1cm}}
 & hydrosimulation & $20\pm 1 \%$ \\
 \hline
 $V_{\rm peak}$   & \pbox{20cm}{\vspace{0.1cm}Maximum value of the velocity in
 a spherically-averaged \\ rotation curve ever achieved by 
 the largest subhalo \vspace{0.1cm}} & hydrosimulation & $20\pm 1 \%$ \\ [1ex]
 \hline
 $V_{\rm max}$  & \pbox{20cm}{\vspace{0.1cm}Maximum value of the velocity in a
 spherically-averaged \\ rotation curve for the largest subhalo \vspace{0.1cm}}
 & hydrosimulation & $20\pm 1$\% \\
 \hline
 $M_{\rm FoF}$  & Sum of the individual masses of every particle in this group & hydrosimulation & $8\pm 1$\% \\
 \hline
\end{tabular}
\caption{Percentage difference between the correlation function
of the galaxies assigned in TNG300-1-Dark using the hydrodynamical simulation outputs and the ``basic'' HOD (bHOD)
 prescription averaged over the scales $r = 1-20$ Mpc$/h$
for different proxies of the host halo mass. The uncertainty we report
comes from the fact that we shuffle the data randomly,
so each realization offers a slightly different shape of the correlation
function. We estimate it as the standard deviation around the mean
for $\sim 10$ random realizations. Fortunately, the scatter is small and
does not change the overall conclusions.}
\label{table:proxy}
\end{table*}

\subsection{Checking box-size effects and cosmic variance with \Abacus{Abacus}}
Some of the main concerns regarding
the robustness of our results
are the limited size of the TNG box and the cosmic variance,
which may play a significant
role on the
scales considered.
To test whether that is the case, we repeat the
procedure from Section \ref{subsec:algo} of shuffling
the halo occupations in 5\%-mass bins, but
this time using a multitude of N-body simulation boxes of 
similar size and resolution to TNG300-1 (27 in total).
To this end,
we select an initial Poisson draw from the
HOD distribution for {each of the 27 boxes} to be the ``true''
galaxy distribution and then apply the
shuffling procedure.
We finally examine the scatter in all
27 boxes in an effort
to better quantify the statistical significance of our
TNG results. The cosmic variance check also helps quantify
whether TNG300 is simply an abnormal region 
of space where the HOD fails as a statistical fluctuation,
rather than a physical effect. 

Here is the recipe in more detail:
\begin{itemize}  
    \item[1.] Numerically derive the average 
    number of galaxies per halo as a function of the halo mass
    measured in logarithmic bins from the TNG300-1 full-physics simulation
    box, i.e. the histogram in Fig. \ref{fig:hod_fof}.
    \item[2.] Make a Poisson draw for each ROCKSTAR or FoF halo 
    in the \Abacus{Abacus} $L_{\rm box} = 720 {\rm Mpc}/h$ box to obtain the 
    number of galaxies it contains. (We use both halo catalogs
    for consistency checking.)
    \item[3.] Evaluate the galaxy number density and if necessary
    renormalize
    the halo masses in \Abacus{Abacus} (multiplying them by
    a $\mathcal{O}(1)$ factor) until the galaxy number densities 
    between TNG300-1-Dark and \Abacus{Abacus} match.
    \item[4.] Split the \Abacus{Abacus} box into 27 cubic subboxes 
    of length $L_{\rm box} = \frac{1}{3} 720 {\rm Mpc}/h = 240 {\rm Mpc}/h$.
    The  size of each subbox is now comparable to that of
    TNG300, $L_{\rm box} = 205 {\rm Mpc}/h$.
    \item[5.] For each of the 27 subboxes, we repeat the steps in
    Section \ref{subsec:algo} with mass proxy $M_{\rm 200m}$, as before.
    The only difference is that within the halo boundaries,
    we place the central galaxy and its satellites inside 
    the dark-matter halo, so that their positions trace the best-fitting
    NFW profile \citep{1996ApJ...462..563N,1997ApJ...490..493N} 
    that quantifies the DM 
    distribution in these halos. This choice does not affect the galaxy
    clustering results on large scales. The subboxes
    do not have periodic boundary conditions, so we 
    compute the correlation function using
    \begin{equation}
        \hat \xi_{\rm LS}(r) = \bigg(\frac{N_{\rm rand}}{N_{\rm 
        data}}\bigg)^2 \frac{DD(r)}{RR(r)} -
        2 \frac{N_{\rm rand}}{N_{\rm data}} \frac{DR(r)}{RR(r)} + 1
    \end{equation}
    with $N_{\rm rand} = 35 \ N_{\rm data}$  random points \citep{1993ApJ...412...64L}.
\end{itemize}

\begin{figure}
\centering  
\includegraphics[width=0.5\textwidth]{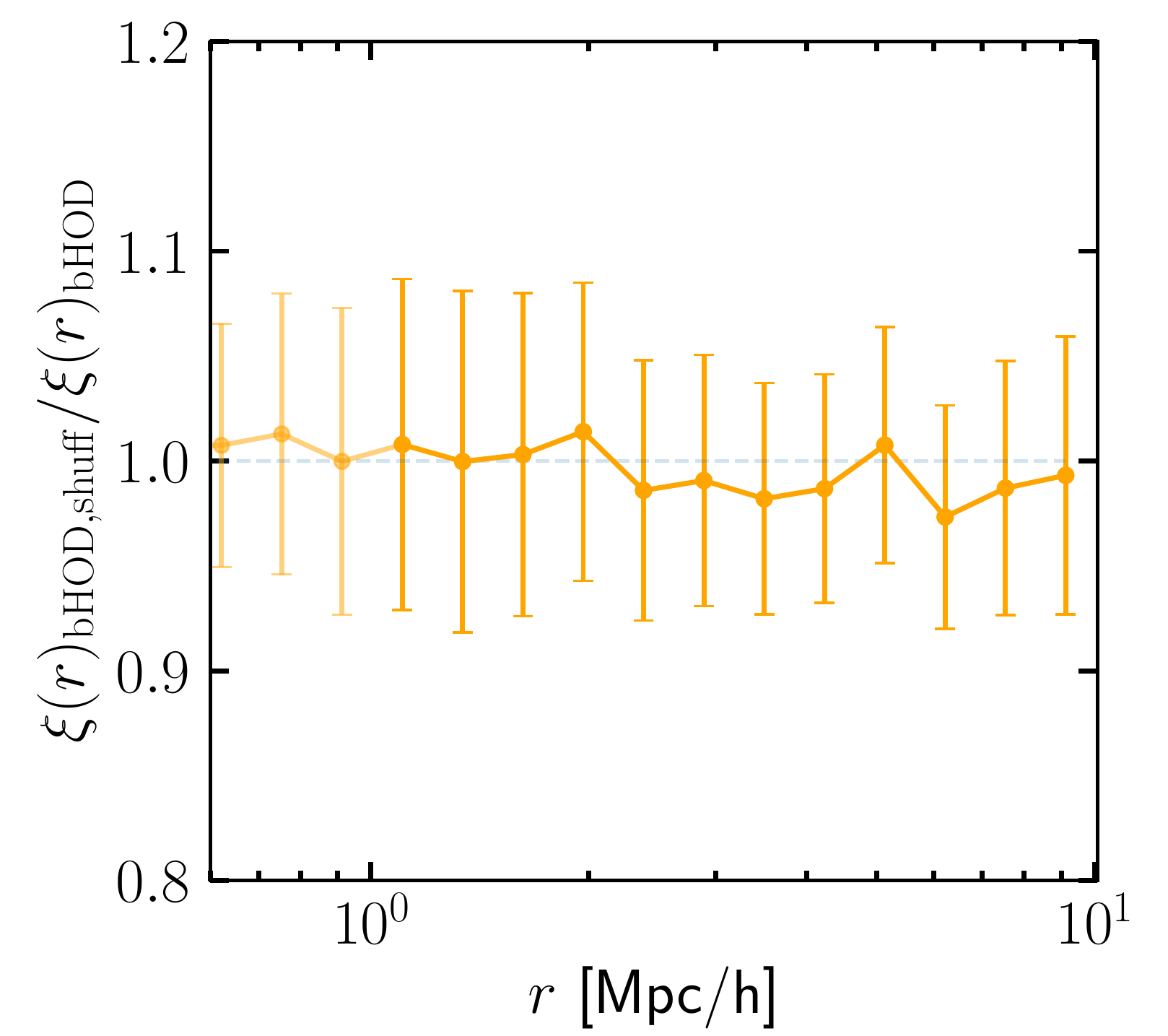}
\caption{Correlation function ratio between 
the randomly shuffled
basic model assigned halo 
occupations (bHOD) and the fiducial (unshuffled)
case in 27 \Abacus{Abacus} subboxes with 
$L_{\rm box} = 240 {\rm Mpc}/h$. In
contrast with the TNG300-1 case where we
see a discrepancy of 15\% 
on large scales, the ratio in the \Abacus{Abacus} 
case is consistent with 1
on all scales ({differs from 1 at 0.3\% 
for $r>$ 1 Mpc/$h$}). Since we
cannot mimic the statistical significance of the
TNG300-1 result with
a simple HOD model, other factors such as assembly bias and
baryonic feedback might be at play in the TNG300-1 case.
The results we obtain using 
the FoF-defined halos do not differ significantly from 
the case of the ROCKSTAR-defined halos.} 
\label{fig:shuff_rat_ab}
\end{figure}

The results are shown in Fig. \ref{fig:shuff_rat_ab}.
We see that the mean ratio curve of all 27
\Abacus{Abacus} {subboxes differs by only 0.3\% from
1 on the scales we are interested in, 
$1 \ {\rm Mpc}/h < r < 10 \ {\rm Mpc}/h$,
which is consistent with what would
be expected if one randomly shuffles Poisson samples 
drawn from the same mean distribution. This
is effectively what we do when we shuffle the galaxy
occupation numbers in 5\% mass bins, as the mean HOD 
is roughly constant for such small mass changes. 
The standard deviation  is 6.0\%,}
which is smaller than the observed average 10-20\% deviation
in the case of TNG300-1 (Table \ref{table:proxy}). This gives us
confidence the TNG result shown in Fig.
\ref{fig:shuff_rat_ab} is not a manifestation of cosmic
variance or box-size effects.

\section{The search for a second parameter}
\label{subsec:sec}
In this section, we explore which secondary properties of the
halo have the strongest effect on the large-scale clustering
of galaxies. Previous analyses have suggested that, in addition
to halo mass, the overdensity in which the halo resides and 
the concentration of its DM density profile may also play 
defining roles in setting the galaxy bias \citep{2018ApJ...853...84Z,2018MNRAS.480.3978A,sownak}.
Other properties such as halo spin and accretion rate history
have been shown to be either of tertiary importance or
strongly correlated with the halo concentration.

\subsection{Candidates for secondary parameter}
{To investigate the extent to which secondary properties 
affect the galaxy clustering on large scales, we run
a simple test. We again
apply the procedure in Section} \ref{subsec:algo}
{to TNG300-1, but
instead of randomly shuffling the halo occupation numbers,
we assign the galaxy number to each halo within the given
5\% mass bin in order of decreasing or increasing value of
the secondary halo property (i.e. 2'b) in the procedure)}.

\subsubsection{Local environment}
{The effect of environment on the halo and galaxy
clustering has been studied extensively in the literature} \citep{2007MNRAS.378..641A,2017A&A...598A.103P,
2018MNRAS.476.5442P,2018MNRAS.473.2486S}.
A halo residing in a dense region is expected to contain
more galaxies on average than a halo in an underdense
region. This is because halos in overdense regions 
experience more mergers, whereas those in underdense 
regions have more mass accreted in the form of smooth material.
To assess the extent to which local environment affects 
the large-scale clustering, we adopt the following
definition for ``environment factor'' for each halo:
\begin{itemize}
\item[1.] Count the number of subhalos (as defined via the SUBFIND
  algorithm
  \citep{Springel:2000qu}) in a radius of
  $R_{\rm tot \ env} = 5 \ {\rm Mpc/h}$
  centered on the halo centre and sum
  up their masses, $M_{\rm tot \ env}$. 
\item[2.] Count the number of subhalos in a radius of
  $R_{\rm 200c}$ and sum up their masses, $M_{\rm 200c \ env}$.
\item[3.] Subtract the mass within the 200 critical radius from the
  total mass contained in the halo environment and obtain the mass of
  the environment, $M_{\rm env} = M_{\rm tot \ env} - M_{\rm 200c \ env}$.
\item[4.] Normalize by the average environment mass of all halos
  in the HOD and define the environment factor,
  $f_{\rm env} \equiv M_{\rm env}/\bar M_{\rm env}$.
\end{itemize}
Ordering the halo occupation number within each 5\% mass bin in order of largest
to smallest environment factor, $f_{\rm env}$, we compute the correlation function
and compare it with the fiducial case. We see a strong bump near
$r \approx 5 {\rm Mpc}/h$ in the 
second top panel of Fig. \ref{fig:all_secondary} {I}
which suggests that environment might play a more crucial
role in determining the clustering of galaxies on
large scales than expected. An important caveat in
the environment parameter definition is that we condition
on exactly the galaxies that will be counted
in the correlation function, i.e. those separated by
$\sim$ 5 Mpc$/h$.
We explore this parameter further in Section \ref{subsec:env}.

\subsubsection{Mass measure assuming
virial theorem}
According to the virial theorem, the mass of a
bound object can be estimated from
\begin{equation}
    \frac{G M_{\rm vir}}{R_{\rm vir}} = \sigma^2 ,
\end{equation}
where $\sigma$ is the velocity dispersion.
Thus, for a virialized structure, the
combination of $\sigma^2 R_{\rm vir}$
may be interpreted as an excellent 
mass proxy. However,
this rests on the assumption that
the FoF halos of TNG300 are virialized
structures and that is unlikely to
apply to all particles belonging to
the edges of a large halo.

The most widely accepted
choice for a virialized mass proxy is
$M_{\rm 200m}$, but there are many other mass
proxies which can be used, none of 
which are perfect. Here we condition on
a second mass proxy, 
the combination $\sigma^2 R_{\rm halfmass}$.
We use the velocity dispersion $\sigma$ of the most
massive subhalo (identified through
SUBFIND) in the given halo,
and the halfmass 
radius $R_{\rm halfmass}$ of again the most
massive subhalo. This quantity could
capture the mass of the
largest bound structure of the halo through
its dynamical behavior. 

The TNG halo catalog reveals a strong correlation
between conventional halo mass proxies such
as $M_{\rm 200m}$ and the combination 
$\sigma^2 R_{\rm halfmass}$. We first attempt
to use this parameter as mass proxy
and apply a random shuffling in 5\% mass bins.
From Table \ref{table:proxy} we learn that
it is not as effective as some of the other mass
proxies (e.g. $M_{\rm 200m}$).
We next condition on $\sigma^2 R_{\rm halfmass}$ as 
a secondary parameter, ordering it in reverse so
that halos with smaller values of $\sigma^2 R_{\rm halfmass}$
become hosts to a larger number of galaxies.
The motivation for doing so is that the
velocity dispersion is expected to have
an inverse relationship with halo occupation at fixed mass \citep{sownak}
and a smaller half-mass radius of the central galaxy
might be indicative of a galaxy cluster where the satellites
have not been accreted onto the central galaxy. 
Intriguingly,
the result in Fig. \ref{fig:all_secondary} {II}
indicates that indeed we get very strong
large-scale correlation when we condition
on this combination of parameters. This
suggests that dynamical descriptions of
the halo are likely to tie more directly to
its merger history and thus to the expected
number of the galaxies it is hosting. A possible
explanation is that similarly to the velocity
dispersion, $\sigma^2 R_{\rm halfmass}$
is related to concentration {(i.e. the central density) of the halo, which is correlated strongly with halo occupancy} (see Section~\ref{sec:concentration}). The additional
effect from $R_{\rm halfmass}$ would be that
the larger the central subhalo (in radius), 
the more likely it is to have consumed
the smaller subhalos surrounding it,
which suggests that there are more satellite
galaxies on average for an object with a 
small value of $\sigma^2 R_{\rm halfmass}$. {Each of these effects can, in turn, shift the clustering relative to the case where halo occupancy is defined by halo mass only.}

Furthermore, we explored a few other related
parameters such as $M_{\rm cent}/R_{\rm halfmass}$
as a measure of the potential depth and
$M_{\rm cent}/(\sigma^2 R_{\rm halfmass})$ as a measure
of the extent to which halos are
virialized, which resulted in a 3\% and a 0.2\% 
increase of the large-scale
correlation function
with respect to the ``basic'' HOD model
(i.e. in the direction of the hydrodynamical simulation),
respectively.

\subsubsection{Velocity anisotropy}
In Jeans' modeling \citep{1987ApJ...313..121M}, 
there is a well-known degeneracy between the
mass profile of a distribution of particles 
and the velocity anisotropy of orbits that trace the resulting potential. 
This is known as the ``mass-anisotropy degeneracy''.
The velocity anisotropy is
defined as \citep{1987gady.book.....B}
\begin{equation}
    \beta = 1-\frac{\sigma_{\rm tan}^2}{2 \sigma_{\rm rad}^2},
\end{equation}
where $\sigma_{\rm tan}$ and $\sigma_{\rm rad}$
are the tangential and radial velocity dispersions, respectively.
We calculate these quantities over all particles in the
FoF halo by projecting the velocity of each particle
along and perpendicular to the radial direction
(defined with respect to the position of the particle with
the minimum gravitational potential energy) and then computing the
standard deviation of each component \citep{2019arXiv190302007R}. 
It is important to realize that $\beta$
depends on the shape of the halo, so, 
similarly to the 3D dispersion, it captures
information from the full phase-space structure of the parent halo. 
The limits of this parameter, $-\infty$ and 1, correspond to
radially and tangentially dominated velocity dispersions,
respectively, while $\beta = 0$ indicates an isotropic
distribution of particle orbits.

\begin{figure*}
\centering  
\includegraphics[width=1\textwidth]{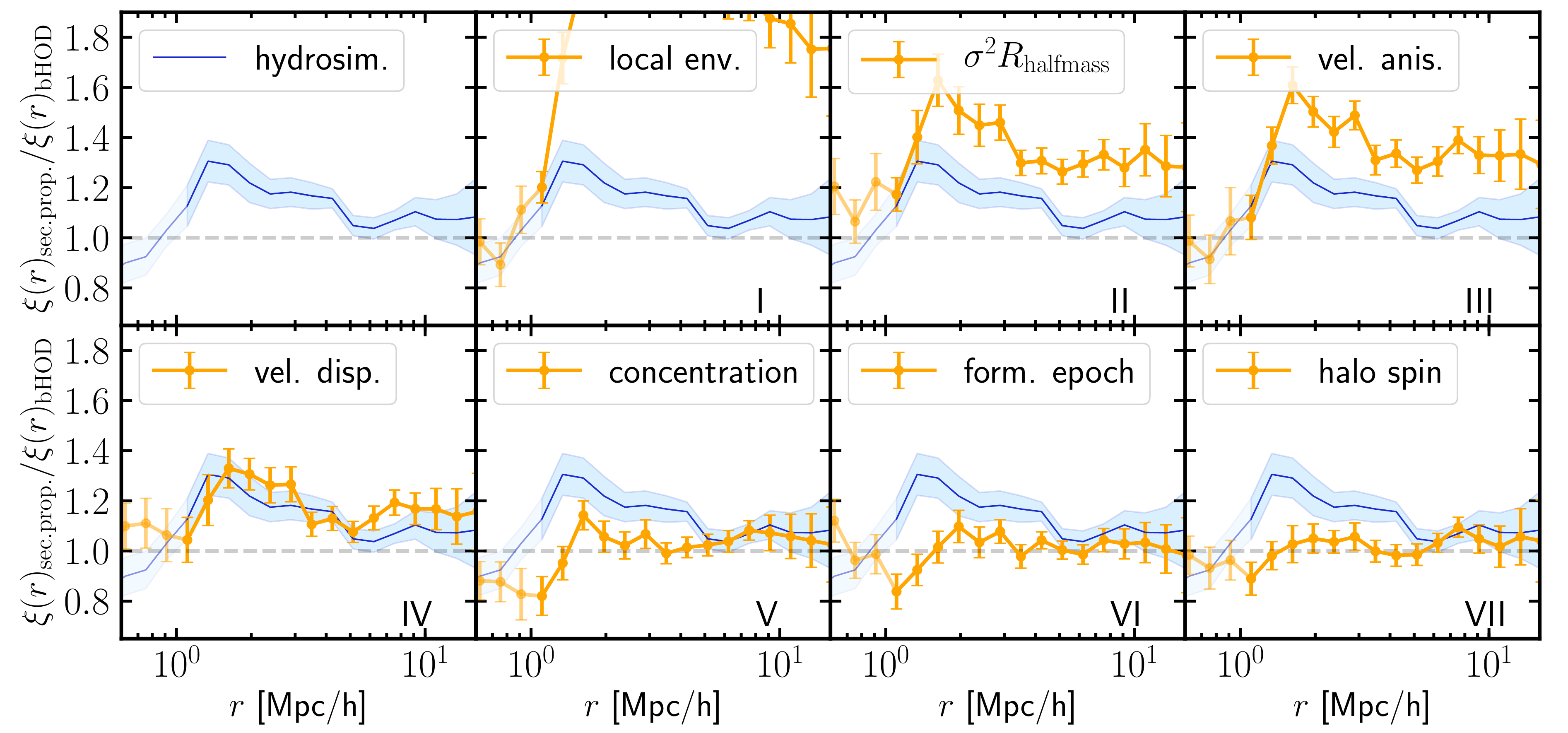}
\caption{Ratio between the correlation function
of the galaxies assigned in TNG300-1-Dark
and those in the ``basic'' HOD (bHOD) prescription. The assignment is
accomplished by conditioning on different secondary halo parameters
and reordering the halo occupation numbers in 5\% mass bins. The
secondary parameters considered are
the environment factor $f_{\rm env}$, the dynamical virial mass
$\sigma^2 R_{\rm halfmass}$, the velocity anisotropy $\beta$,
the velocity dispersion $\sigma$, halo concentration $c$,
the formation epoch, and the halo spin $\lambda$, always shown in orange
(\textit{left to right, top to bottom}).
In shaded blue
we show the result from the bijective match between
the halos in the full-physics TNG300-1 box and the dark-matter-only TNG300-1 box.
The mass proxy used in all cases is $M_{\rm 200m}$.} 
\label{fig:all_secondary}
\end{figure*}

{Previous works have shown that halos with a high value of $\beta$ cluster more weakly than those with low $\beta$}
\citep{2010ApJ...708..469F,2019arXiv190302007R}. For our
reordering test, we {therefore} assign higher number of galaxies
to halos with smaller dispersion anisotropy, analogously
to how we treat next the velocity dispersion as a secondary parameter.
The {net} effect of this can be seen in Fig. \ref{fig:all_secondary} {III}, which
shows a significant increase of the galaxy clustering 
when comparing it with the true galaxy
clustering on large scales. If we quantify this difference,
we see that on average this result overshoots the ``basic'' HOD by about 36\%  
on large scales. {In our parameter search, the velocity anisotropy, $\beta$} is the second most influential
{secondary} parameter on the clustering of galaxies, after local
environment (see Table \ref{table:second}).

One plausible explanation for the more 
isotropic velocity distribution (low
value of $\beta$) of the more clustered
halos (and galaxies) is that the 
impact parameters of the merging subhalos
are larger due to 
deflections caused by gravity 
shortly before accretion,
hence $\sigma_{\rm tan}$ acquires
a larger value. 
Since mergers are what 
dominates the {mass growth of halos} in high density {(i.e. more clustered)} regions
\citep{2009MNRAS.394.1825F,2010MNRAS.401.2245F},
they may be influential in determining the velocity 
structure of these halos and are thus
closely related to the number of subhalos
(and galaxies) residing in them. 
As for lower density regions,
accretion occurs in a 
more radial fashion since the 
gravitational field is dominated
by the largest subhalo. This
leaves an imprint on the velocity
structure of the halo and the 
value of $\beta$ increases
\citep{2009MNRAS.394.1825F,2010ApJ...708..469F}. 

\subsubsection{Velocity Dispersion}
Similarly to halo concentration ({discussed next}), the velocity
dispersion $\sigma$, is correlated
with accretion history. However, contrary to the halo concentration,
it might be a more directly
correlated measure of the most recent merger history of the halo
as it uses dynamical information {regarding particle velocities}
rather than {simply their positions}.
{Using the TNG simulations, it has been shown that}
velocity dispersion has an inverse relationship
with halo occupation at fixed mass \citep{sownak}. For this reason,
when applying the reordering procedure outlined in
Section \ref{subsec:algo}, we give highest priority
to the halos with smallest dispersion velocity.
We present this result in Fig. \ref{fig:all_secondary} {IV},
which shows a moderate increase of the clustering of HOD 
galaxies ordered in terms of their velocity
dispersion when comparing it with the true galaxy
clustering on large scales. If we quantify this difference,
we see that on average this result overshoots the ``basic'' HOD by 18\%  
on large scales. This is the fourth most influential {secondary}
parameter on the clustering of galaxies 
(see Table \ref{table:second}).

\subsubsection{Halo concentration}
\label{sec:concentration}
The concentration of the halo is closely related to its 
accretion and formation history and has been
well studied in the literature
\citep{1997ApJ...490..493N,Wechsler:2001cs,2014MNRAS.441..378L,2016MNRAS.460.1214L}. 
In simplified terms,
the larger the number of recent 
mergers it has undergone, the more spatially spread out
 its subhalos are likely to be; i.e. the smaller its 
 concentration will be \citep{sownak}.
A larger number of satellites might imply
 that halos of smaller concentration have more highly 
clustered galaxies than the more compact ones. Another
 important consideration, however, is that as we get to 
smaller halo masses,  where on average we expect 1 
or 0 galaxies (i.e. only a central), more concentrated 
halos are more likely to host a galaxy potentially because
 the gravitational well is deeper, so gas is more likely 
to collapse towards the center and form stars. 
In addition, its dependence on halo mass and
environment has also been thoroughly explored
\citep{2001MNRAS.321..559B,2014MNRAS.441..378L,2015ApJ...799..108D}.
We include it
as a likely candidate for assembly bias affecting galaxy clustering
and employ the following proxy for halo concentration, 
\begin{equation}
c = R_{\rm 200c}/R_{\rm max},
\end{equation}
where $R_{\rm max}$ is the radius at which the maximum circular velocity
is attained.

We implement the dependence on halo concentration by splitting the halos in
two groups depending on their mass: if their mass is larger than {some threshold value,}
 $M^\ast$, which we define as $\langle N_{\rm gal}(M^\ast) \rangle = 1$,
 we reorder the halo occupation numbers within each 5\% mass bin,
starting with the least concentrated halos. If, however, the halos within
the mass bin have masses smaller than  $M^\ast$,
then we give priority, i.e. grant a central
galaxy, to the more concentrated ones.
The result is shown in Table \ref{table:second} 
and Fig. \ref{fig:all_secondary} {V}. Although
we see an improvement of about 3\% compared with the 
random shuffling case, the galaxy clustering on large scales
is still not recovered at a sufficient precision for future experiments
and a 9\% discrepancy remains. {This suggests that halo concentration, on its own, is not an effective secondary parameter with which to augment the ``basic'' HOD model, being less influential than environment and velocity anisotropy in altering the large-scale clustering of galaxies.}

\begin{table*}
\centering
\begin{tabular}{||c | c| c| c||} 
 \hline
 Mass proxy & Secondary property & Secondary property definition used in bHOD & Difference from bHOD   \\ [0.5ex] 
 \hline\hline
 $M_{\rm 200m}$  & 
 hydrosimulation & \pbox{20cm}{\vspace{0.1cm}
   Hydrosimulation results from TNG300-1 for \\
 the galaxy distribution \vspace{0.1cm}} & $15\pm 1 \%$ \\
 \hline
  $M_{\rm 200m}$ & local environment & \pbox{20cm}{\vspace{0.1cm}$f_{\rm env}$, mean density in an annulus of $R_{\rm 200m}$ \\ to $R_{\rm tot \ env} =$ 5 Mpc$/h$ surrounding the halo \vspace{0.1cm}} & $98.6 \%$ \\
 \hline
 $M_{\rm 200m}$ & $\sigma^2 R_{\rm halfmass}$ & \pbox{20cm}{\vspace{0.1cm}Dispersion velocity times the radius containing \\ half of the total mass of the largest subhalo \vspace{0.1cm}}  & $35.4 \%$ \\
 \hline
  $M_{\rm 200m}$  & velocity anisotropy & \pbox{20cm}{\vspace{0.1cm}$\beta =1-\sigma^2_{\rm rad}/2 \sigma^2_{\rm tan}$ of the largest subhalo  ($\sigma_{\rm rad}$, \\ $\sigma_{\rm rad}$ are tangential and radial dispersion) \vspace{0.1cm}} & $35.8$\% \\
 \hline
 $M_{\rm 200m}$  & dispersion velocity & \pbox{20cm}{\vspace{0.1cm}$\sigma$, one-dimensional velocity dispersion of all \\ the member particles of the largest subhalo \vspace{0.1cm}} & $17.9$\% \\
 \hline
  $M_{\rm 200m}$ & $M_{\rm cent}/R_{\rm halfmass}$ & \pbox{20cm}{\vspace{0.1cm}Mass of the largest subhalo divided by the \\ radius containing half of its total mass \vspace{0.1cm}} & $6.1 \%$ \\
 \hline
  $M_{\rm 200m}$ & halo concentration & \pbox{20cm}{\vspace{0.1cm}$c = R_{\rm 200c}/R_{\rm max}$ ($R_{\rm max}$ is the comoving radius \\ where $V_{\rm max}$ of the largest subhalo is achieved) \vspace{0.1cm}} & $2.7 \%$ \\
  \hline
 $M_{\rm 200m}$ & halo spin & \pbox{20cm}{\vspace{0.1cm}$\lambda = J_{\rm cent}/{\sqrt2 M_{\rm 200m} R_{\rm 200m} V_{\rm 200m}}$ ($J_{\rm cent}$ is total \\ angular momentum of the largest subhalo) \vspace{0.1cm}} & $2.0 \%$ \\
 \hline
 $M_{\rm 200m}$ & formation epoch & \pbox{20cm}{\vspace{0.1cm}Snapshot during which the largest subhalo \\ acquired half of its total present mass \vspace{0.1cm}} & $0.6 \%$ \\
  \hline
\end{tabular}
\caption{Percentage difference between the correlation function
of the galaxies assigned in TNG300-1-Dark using the hydrodynamical simulation
outputs conditioned on different secondary halo parameters
and the ``basic'' HOD (bHOD)
prescription averaged over the scales 
$r = 1-20$ Mpc$/h$. The most influential assembly bias
parameters seem to be environment followed by
the virial-mass-like combination $\sigma^2 R_{\rm halfmass}$ and the
halo velocity anisotropy.}
\label{table:second}
\end{table*}

\subsubsection{Formation Epoch}
{The characteristic formation epoch of a halo is a direct indicator of its past accretion history at fixed present-day mass.}
{Here, we define formation epoch} 
as the epoch at which the
halo has acquired 50\% of its
present-day mass (using the TNG merger trees
\citep{2019ComAC...6....2N}), early-forming halos are expected to
have had more time to accrete galaxies
due to an expected larger number of
mergers and also have, on
average, deeper potentials 
(i.e. early-forming halos are more concentrated). 
However, the formation epoch {(as it is defined here)}
informs us about the
ancient history of the halo and
is not as sensitive to the most recent
merger events it has undergone, nor is
it necessarily correlated with the
age of the galaxies residing within it.
Even so, many of the early-formed
galaxies have consumed their satellites,
so modeling the relationship between
occupation number and epoch of formation
is not an easy task.
This is in fact what we find
in Table \ref{table:second} and Fig. 
\ref{fig:all_secondary} {VI}. Formation
history, defined in the way described above,
seems to be the weakest assembly bias candidate.
It is possible that a definition of formation
epoch, spanning over a longer time range
or reflective of more recent merger events, might 
have a more substantial effect on galaxy clustering
on the scales of our interest. {In particular, a definition that captures both ancient and more recent formation events may be more influential as a secondary parameter with which to augment the ``basic'' HOD; we leave the exploration of such a metric to future work.}

\subsubsection{Spin}
{The final secondary parameter we explore in this section is halo spin.} This provides a measure of the angular
momentum acquired by the halo and its
dependence on other halo properties
has been well studied 
\citep{2001ApJ...555..240B,2007MNRAS.376..215B,2016MNRAS.462..893R,2019MNRAS.486.1156J}.
The measurement of this parameter turns
out to be quite sensitive to the particle resolution
\citep[the smaller the number of particles in a halo, the
larger the error,][]{2014MNRAS.437.1894O,2017MNRAS.471.2871B}.

We adopt the following
definition of dimensionless
spin $\lambda$ proposed by
 \citet{2001ApJ...555..240B}
 \begin{equation}
     \lambda = \frac{J_{\rm vir}}{\sqrt2 M_{\rm vir} R_{\rm vir} V_{\rm vir}},
 \end{equation}
where $J_{\rm vir}$ is
the angular momentum inside
a sphere of radius $R_{\rm vir}$ of mass $M_{\rm vir}$
and with halo circular
velocity $V_{\rm vir} = \sqrt{G M_{\rm vir}/R_{\rm vir}}$.
Since we only consider subhalos comprised of $\geq 10000$
star particles as galaxies, the sample of halos for which
we compute the spin are predominantly $\log M \gtrsim 12.7$,
and are sufficiently well-resolved so that noisy spin
measurements are not an issue \citep{2017MNRAS.471.2871B}.

In Table \ref{table:second} and Fig. \ref{fig:all_secondary}
{VII}, we show the resulting 
percentage difference. It shows a moderate improvement
of about 2\% with respect to the randomly shuffled case {when
we preferentially give the largest number of galaxies to
the halos with largest spin (or in this case, $\lambda$)}.
Hence, we can conclude from this analysis that 
halo spin plays a minor role in
predicting the occupation number of a halo.

\subsection{Predicting the correlation function shape}
\label{subsec:corr}
In Section \ref{subsec:sec}, we showed that the parameters
that are most influential in shifting the large-scale clustering
in the direction of the hydrosimulation result are the local 
environment parameter $f_{\rm env}$,                  
the velocity anisotropy                        
$\beta$, and the dynamical mass proxy $\sigma^2 R_{\rm halfmass}$. 
We did this by a perfect association of ranks between 
the halo occupation and the second parameter {(e.g. the halo with
the largest value of $f_{\rm env}$ gets assigned the largest number 
of galaxies in each 5\% mass bin)}. If one wants 
to consider an imperfect association which recovers
the clustering of galaxies in TNG, one needs to introduce a new
procedure that preserves the original distribution
of the occupancies. 

\begin{figure*}
\centering  
\includegraphics[width=1\textwidth]{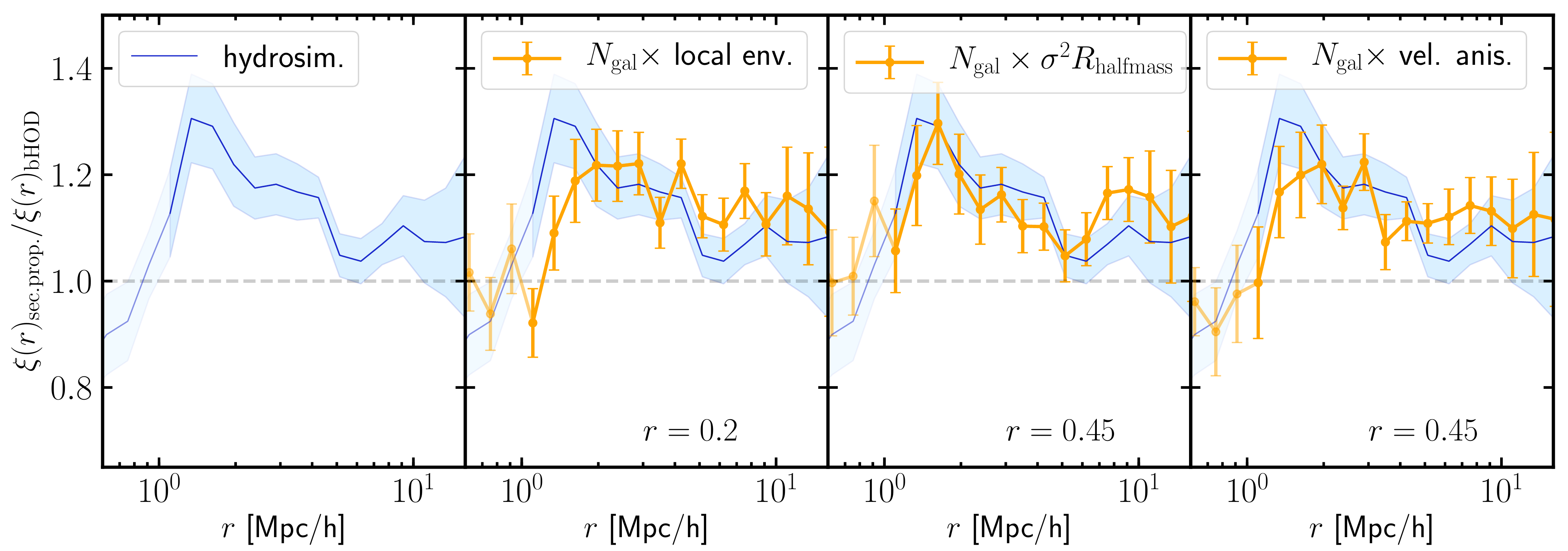}
\caption{Ratio between the correlation function
of the galaxies assigned in TNG300-1-Dark
and those in the ``basic'' HOD (bHOD)
prescription. 
The halo occupation numbers are reordered in 5\% mass bins
according to the strength of the co-dependence
between a given secondary halo parameter and the number of
galaxies hosted by the host halo. The recipe is specified in 
Section \ref{subsec:corr}. The
secondary parameters considered are
the environment factor $f_{\rm env}$, the dynamical virial mass
$\sigma^2 R_{\rm halfmass}$, and the velocity anisotropy $\beta$,
always shown in orange (\textit{left to right})
with the following correlation parameters $r$: $r=0.2$,
$r=0.45$ and $r=0.45$, respectively. 
{The $r$ values are chosen so as to minimize the
fractional difference of  the galaxy correlation function
with respect to the hydrosimulation on large 
scales ($1 \ {\rm Mpc}/h < r < 20 \ {\rm Mpc}/h$).}
In shaded blue
we show the result from the bijective match between
the halos in the full-physics TNG300-1 box and the dark-matter-only TNG300-1 box.
The mass proxy used in all cases is $M_{\rm 200m}$.} 
\label{fig:all_corr}
\end{figure*}

In the spirit of our random-shuffling approach, we apply
the following procedure which attempts to quantify the strength of this
correlation. For each 5\%-mass bin with $N_{\rm h}$ halos in it:  
\begin{itemize}                                                             
  \item[1.] We choose a correlation parameter $r$ between 0 and 1 and draw 
  $N_{\rm h}$ pairs of $(x,y)$ values from a joint Gaussian distribution
  with mean $(0,0)$ and covariance matrix $[[1,r],[r,1]]$. 
  \item[2.] We convert the $(x,y)$ array into an array 
  of integers by ordering 
  the $x$-values from largest to smallest, each one getting a  
  number from 0 to $N_{\rm h}-1$, respectively. We repeat this for  
  the $y$-values, obtaining $N_{\rm h}$ pairs of integers, 
  $(x,y) \rightarrow (i,j)$.
  \item[3.] We now form an array of $N_{\rm h}$, each entry of which
  containing the number of galaxies hosted by a halo, 
  $N_{\rm gal}$, and order it from largest to smallest.
  \item[4.] We convert them to integers  
  $i_{\rm par.}$ between 0 and $N_{\rm h}-1$ as before. 
  Similarly, we create another array filled with
  the values of whichever secondary parameter we are exploring 
  -- $f_{\rm env}$, 
   $\sigma^2 R_{\rm halfmass}$, or  $\beta$, and again convert them
   it into an integer array $j_{\rm par.}$.
   \item[5.] Order the $(i,j)$ pairs in order of the $i$ values 
   and do the same for the array of $i_{\rm par.}$ values.  
   Identifying the $i$'s with the $i_{\rm par.}$'s, we now know what
   the corresponding $j$ value is for each $i_{\rm par.}$. 
   \item[6.] Find the original value of the $j_{\rm par.}^{\rm th}$
   parameter in the secondary property array for each element $i$
   (or equivalently $i_{\rm par.}$) in the pair $(i,j)$, 
   for which $j = j_{\rm par.}$.
   We thus effectively end up with a uniform distribution 
   of discrete correlated pairs, e.g. a Gaussian drawn
   $(N_{\rm gal},f_{\rm env})$ distribution,
   for a given choice of the correlation parameter $r$. 
\end{itemize} 

The amount of  correlation, $r$, between the number
of galaxies per halo and one of the three parameters,
$f_{\rm env}$, $\beta$, and  $\sigma^2 R_{\rm halfmass}$,
which is required to obtain approximately
the same behavior of the correlation function on large
scales for each of the three parameters as in the
hydrosimulation, is
$r = 0.2$, $r = 0.45$ and $r = 0.45$, respectively.
In Fig. \ref{fig:all_corr}, we show what the resulting
correlation functions look like in adopting these values
for the correlation parameter $r$.

\subsection{The environment factor}
\label{subsec:env}
As was shown in Table \ref{table:second}
and Fig. \ref{fig:all_secondary} {I}, the biggest impact on the
clustering of galaxies on large scales after halo mass
is the environment. In this section, we explore
other, more rigorous definitions of the
environment parameter than the one
used previously -- in particular,
we quantify a halo's tidal environment, 
and also split halos based on percentiles of local overdensity.

{The reason we explore both is to test whether
the galaxy distribution is affected specifically by
the tidal environment or whether the conclusions
drawn in that case hold true in the case of
splitting galaxies into different density regions as well.}

\subsubsection{Tidal environment}
\label{subsec:tidal}
To characterize the ``cosmic web'' distribution in
our simulation, we follow the conventional tidal environment
assignment algorithm \citep{1970Ap......6..320D,2007MNRAS.375..489H,2009MNRAS.396.1815F}. First, we evaluate the density field,
$\delta (\mathbf{x})$ using cloud-in-cell (CIC) interpolation on a 
$256^3$ cubic lattice of only the dark matter particles.
We then solve the Poisson equation 
$\nabla^2 \psi = \delta$ and obtain the second derivative 
$\psi_{ij} \equiv \partial^2 \psi / \partial x_i \partial x_j$
in Fourier space, applying a Gaussian smoothing kernel.
Finally, we compute the eigenvalues
$\lambda_1 \leq \lambda_2 \leq \lambda_3$ and define the 4
standard types of environment:
\begin{itemize}
\item \textbf{peaks}: all eigenvalues below the threshold ($\lambda_{\rm th} \geq \lambda_1$)
\item \textbf{filaments}:  one eigenvalue below the threshold ($\lambda_1 \geq \lambda_{\rm th} \geq \lambda_2$)
    \item \textbf{sheets}:  two eigenvalues below the threshold ($\lambda_2 \geq \lambda_{\rm th} \geq \lambda_3$)
    \item \textbf{voids}: all eigenvalues above the threshold ($\lambda_3 \geq \lambda_{\rm th}$).
\end{itemize}
The choice of the threshold value for 
the eigenvalues is somewhat
arbitrary. Here, we pick $\lambda_{\rm th} = 1.2$ in order to
maintain a roughly equal number of galaxies in the 
first three tidal environment types. A given halo is said
to belong in one of four environment types depending on the
environment type of the cell it is located in.

\begin{figure}
\centering  
\includegraphics[width=0.5\textwidth]{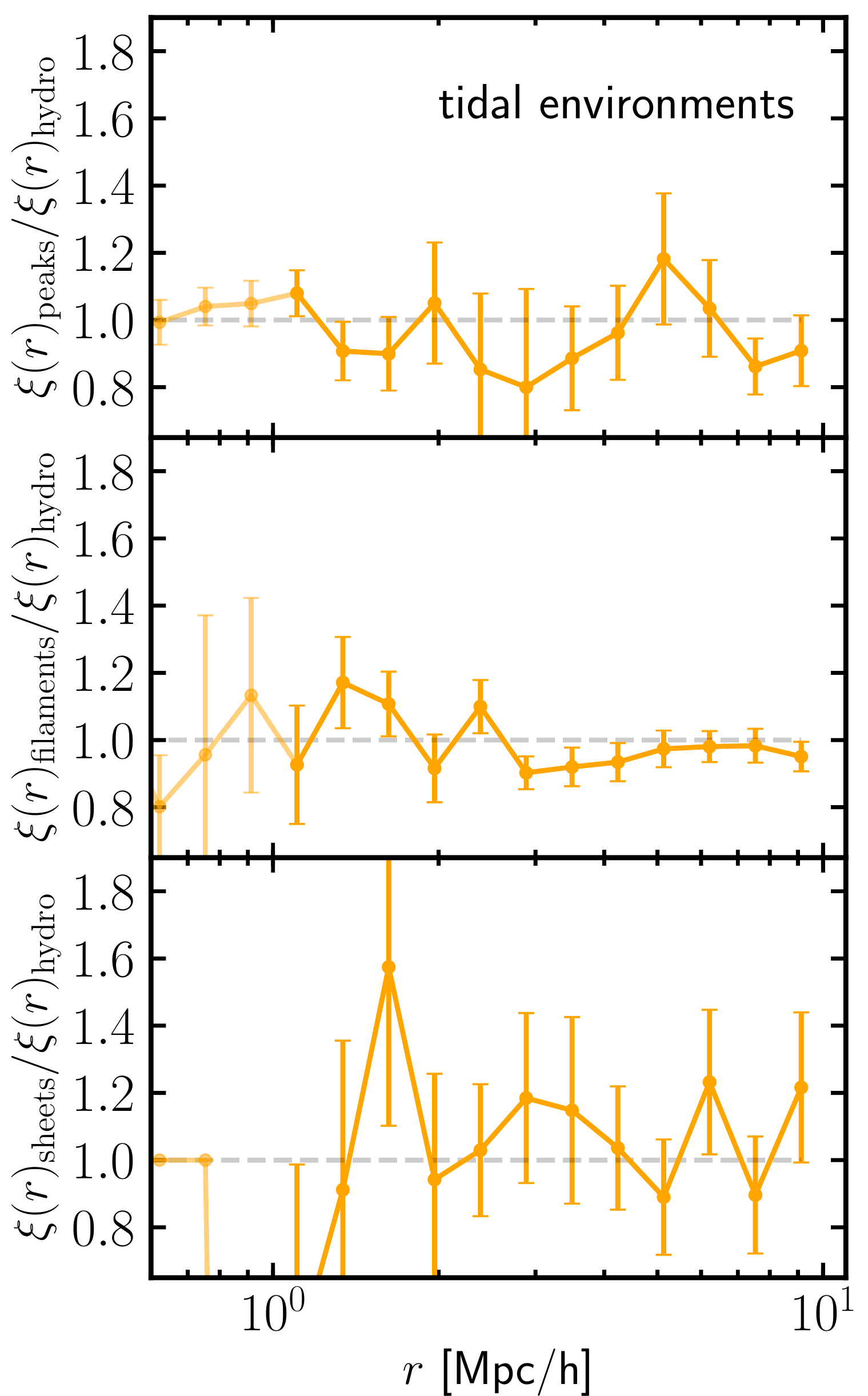}
\caption{Correlation function ratio
between 
the shuffled halo occupations in 5\%
mass bins and the fiducial (unshuffled) case
for halos residing in the peaks
\textit{upper panel},
filaments \textit{middle panel},
and sheets \textit{lower panel}. This result  exhibits
a more modest discrepancy from the \textit{hydrosimulation} 
(\.*) result
compared with what was observed in the
full sample (Fig. \ref{fig:shuff_rat}) for $r>$ 1 Mpc/$h$, 
where the difference between
the hydrosimulation result and the ``basic'' 
HOD one is about 15\%. There is roughly
an equal number of galaxies residing in each of the three
environments. This suggests that galaxies in halos of the
same environment cluster similarly. The error bars
are relatively large both because of the few objects and
the limited volume occupied by the sheets and
knots, in particular.}
\label{fig:shuf_rat_cw}
\end{figure}

We apply the random shuffling procedure in Section
\ref{subsec:algo} to each of the first three regions
defined above (as the last region, the voids, 
turns out to have
0 galaxies for our choice of $\lambda_{\rm th}$
and galaxy definition). In Fig. \ref{fig:shuf_rat_cw},
we show the results of this random shuffling in
each region in 5\% mass bins. We see that the 
galaxy clustering ratio on large scales when we fix the 
environment type and shuffle only within it
is more consistent with full-physics TNG300-1
({differing from it by 4.8\%, 1.1\% and 4.0\% 
in each of the three panels, respectively, for $r>$ 1 Mpc/$h$}).
In our fiducial shuffling plot of all halos
irrespective of their environments, i.e. Fig.
\ref{fig:shuff_rat}, we saw a 15\% difference, which implies
that galaxies belonging to halos in one environment
are likely resorted into halos residing 
in another tidal region.  This suggests that for future models of galaxy
occupation a split
into environment regions might lead
to better agreement between the mock galaxy catalogs
and the true galaxy distribution.

\begin{figure}
\centering  
\includegraphics[width=0.5\textwidth]{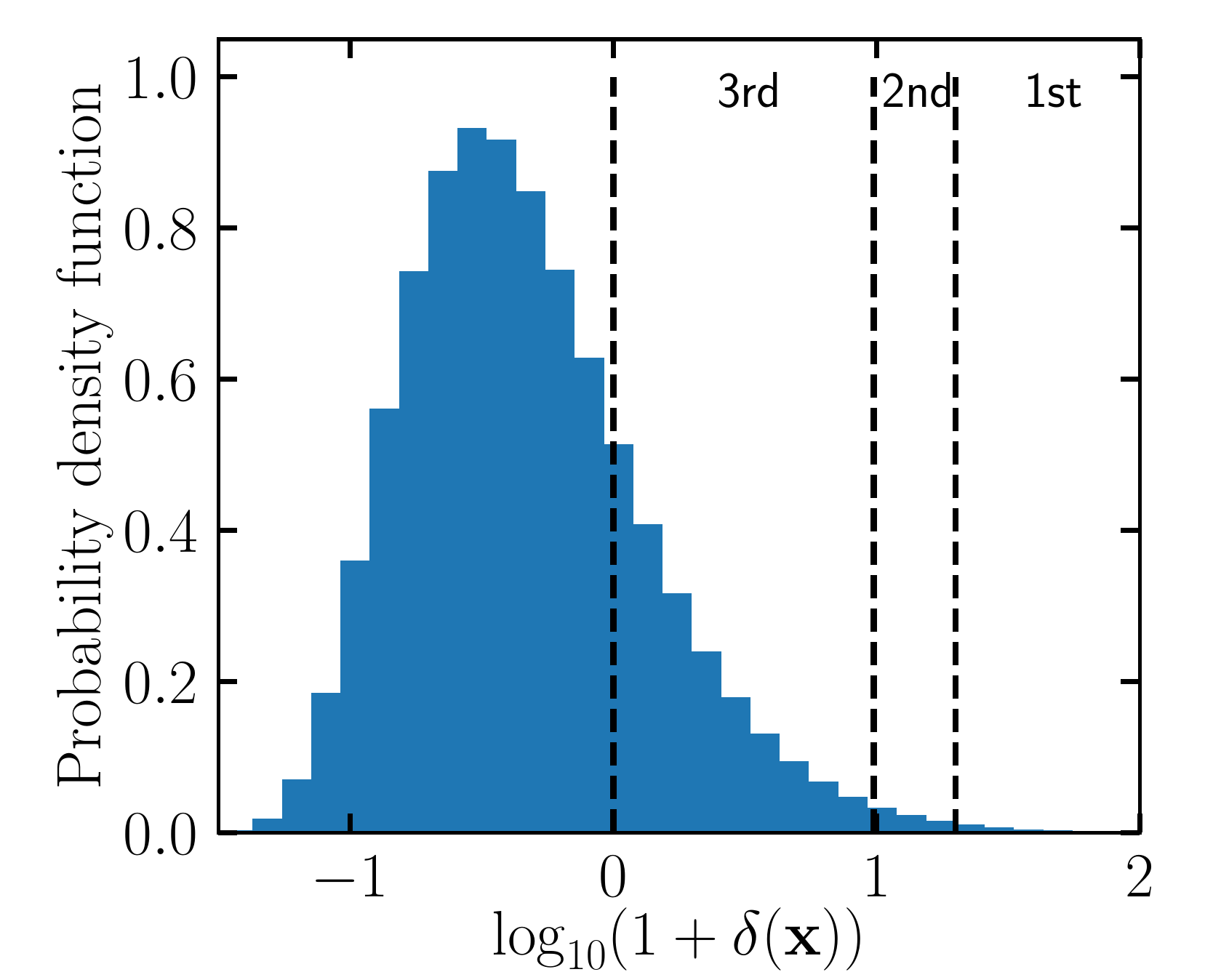}
\caption{Histogram of the flattened 3D-logarithmic
  DM density in TNG300. The splits between the 3 regions are chosen
  such that each one contains a roughly equal number of
  galaxies as would be ascribed from the bijective matches with
  the hydrodynamical simulation.}
\label{fig:log_den}
\end{figure}

\subsubsection{Density percentiles}
We now attempt yet another environment-based reshuffling
scheme, this time parameterizing environment by
$\texttt{d}(\mathbf{x}) \equiv \log_{10}(1+\delta(\mathbf{x}))$, 
where $\delta$ is the local overdensity computed in 
\ref{subsec:tidal}.
We adopt the following ranges for the densest,
second densest and third densest regions, respectively:
$\texttt{d} \in [0,0.99)$, $\texttt{d} \in [0.99,1.3)$,
and $\texttt{d} \in [1.3,3)$, \.*.
We have investigated the
effect of random shuffling in 5\% mass bins
within each region.
The 3D map of the DM density 
of TNG300 is obtained with a CIC assignment
and pixel size of 
$205/256 \ {\rm Mpc}/h \approx 0.8 \ {\rm Mpc}/h$.
\citep{forero}.

\begin{figure}
\centering  
\includegraphics[width=0.5\textwidth]{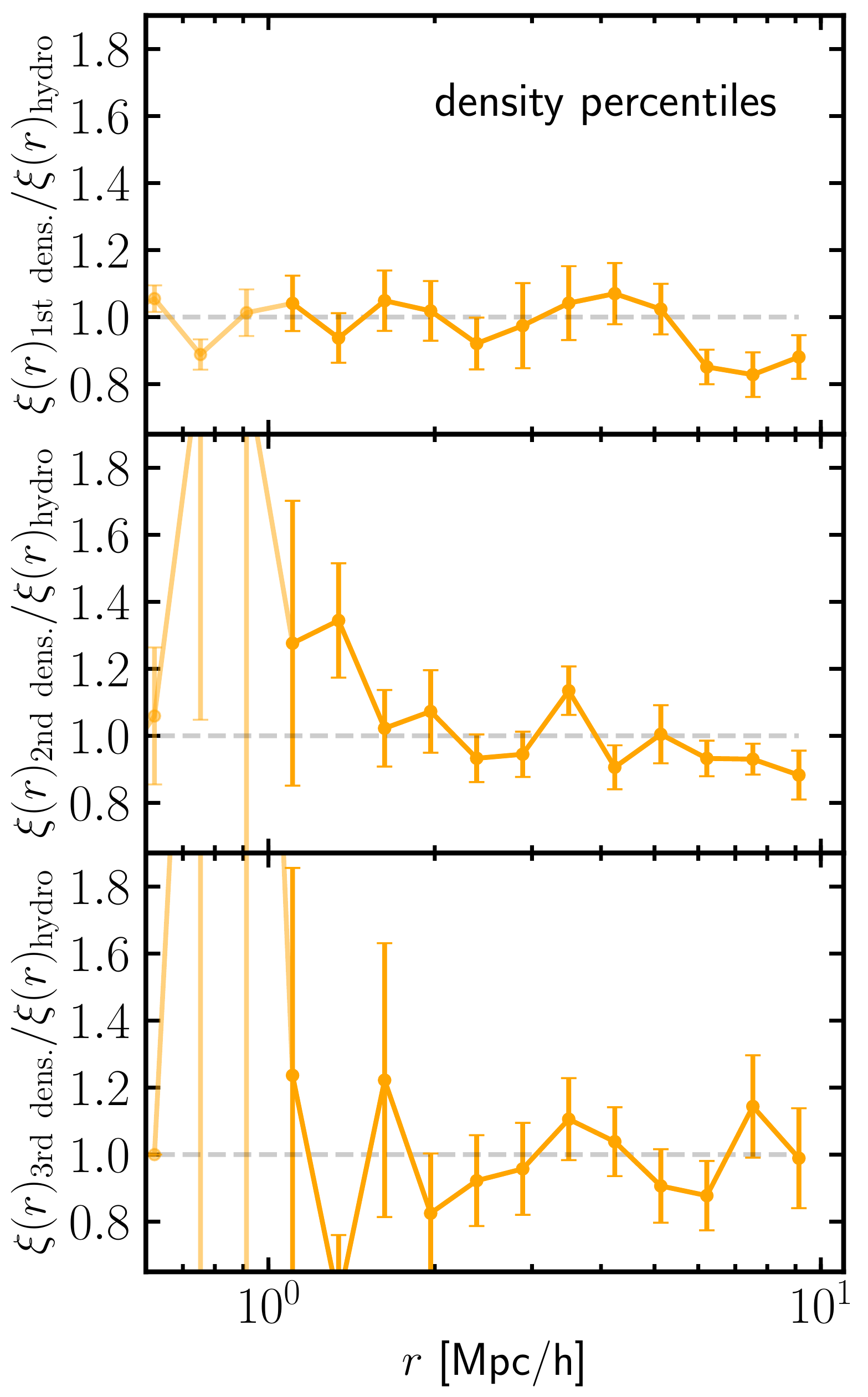}
\caption{Correlation function ratio
between 
the shuffled halo occupations in 5\%
mass bins and the fiducial (unshuffled) case
for halos residing in the densest
(\textit{upper panel}),
second densest (\textit{middle panel}),
and third densest (\textit{lower panel})
regions. There is roughly
an equal number of galaxies residing in each of the three
environments.
This result ({a difference of 3.0\%, 3.2\% 
and 1.8\%, respectively}) does not exhibit
a discrepancy at the level observed in the
full sample (Fig. \ref{fig:shuff_rat}) (15\%)
for $r >$ 1 Mpc/$h$,
which indicates that effects such
as the environment density are vital for determining
the galaxy occupation numbers of halos since
halos in similar environments cluster similarly.}
\label{fig:shuf_rat_den}
\end{figure}

Similarly to the previous case of classifying
and shuffling within each tidal environment region,
we will do the same for each density region, but
first we need to pick bounds which define the density
regions. As in the previous case, we will choose
the boundary points arbitrarily, defined so as to
preserve a roughly constant number of galaxies 
within each region (except the lowest density one). We
then apply the methodology from Section \ref{subsec:algo}.
The results are shown in
Fig. \ref{fig:shuf_rat_den}. The ratio is again
nearly 1 which is far from
the 15\% discrepancy for $r>$ 1 Mpc/$h$
observed in the initial test
we perform on TNG300 ({differing from it by
3.0\%, 3.2\% and 1.8\%  in each of
the three panels, respectively}).
We see a smaller scatter compared with
the cosmic web definition 
\ref{fig:shuf_rat_cw}, which is most likely because
the volume which each of the density regions occupies
is slightly less strictly defined than in the tidal environment
case.

\section{Conclusions}
\label{sec:conc}
The standard galaxy-halo modeling based on the HOD 
methodology predicts
that the number of galaxies
residing in a halo is determined solely by its mass.
Recent findings
have challenged the assumptions of this model \citep{2019arXiv190705424Z},
hinting at discrepancies on large scales, e.g.
when comparing the HOD-inferred result with
weak lensing data \citep{2017MNRAS.467.3024L}.
In this paper, we put the
standard HOD method to a test
by generating a standard HOD galaxy 
sample and comparing it to the ``true'' galaxy
distribution within the same hydrodynamical 
simulation, TNG300-1. By computing 
the correlation function
of galaxies in the HOD catalog with those in 
the ``true'' case, we have shown (Fig. \ref{fig:shuff_rat}) that
the simplest HOD model potentially leads to 
differences of order 15\% at large scales 
($> 1$ Mpc$/h$). 

Before considering physical explanations
for the observed large-scale
discrepancy, we further tested
the statistical rigor of our results. First,
we have repeated
the exercise on 27 N-body \Abacus{Abacus} boxes 
($L_{\rm subbox} = 240 {\rm \ Mpc}/h$)
\citep{abacus},
but this time with purely HOD-assigned 
galaxies (see Fig. \ref{fig:shuff_rat_ab}). 
The fact that Fig. \ref{fig:shuff_rat_ab}
does not exhibit the same level of discrepancy
as Fig. \ref{fig:shuff_rat} implies that
the observed 15\% difference in the TNG simulation box 
is highly unlikely
to be solely an artifact of the limited simulation volume. 
Another possible source of error may be in 
the choice of halo finder.
TNG uses the group-finding technique FoF
\citep{1985ApJ...292..371D}, which {is typically
preferred for on-the-fly implementations over 
alternative algorithms especially in high-resolution
simulations, where computation time plays an important
role}. However, FoF is thought
to be less accurate at finding halos \citep{2009ApJ...692..217L,2011ApJS..195....4M,2011MNRAS.415.2293K}, so we have instead 
run the alternative group-finder
ROCKSTAR \citep{rockstar}, which,
although slower, uses full phase-space information.
This did not alleviate the issue at hand.
We have also experimented (see Table \ref{table:proxy}) with constructing
the HOD using different halo mass proxies
(such as $M_{\rm 200m}$, $M_{\rm 200c}$,
and $V_{\rm peak}$), and find that $M_{\rm 200m}$
serves to minimise the large-scale differences.
However, this is only at the 15\% level,
and does not diminish
the differences in Fig. \ref{fig:shuff_rat} entirely.

This 15\% deviation could be reflective of an
inherent issue with these models. The relationship
between the galaxy and its parent halo mass
is strong, but galaxies
are also known to be biased tracers of the
halo and total mass distributions,
an effect known as ``galaxy assembly bias''.
While properties such as
halo formation time, environment, 
concentration, triaxiality, spin, and 
velocity dispersion play a 
non-negligible role in
the halo clustering,
the relationship 
between the halo properties
and that of the galaxy is not
well understood. In this paper, we
test {8 halo properties beyond mass} --- 
local environment, velocity anisotropy,
$\sigma^2 R_{\rm halfmass}$,
velocity dispersion, halo spin, halo
concentration, depth of the potential, and
formation epoch ---
in an attempt to reveal which ones,
if any, have a direct impact on
the large-scale clustering of galaxies (see Table
\ref{table:second}).

We find that halo environment
correlates very strongly with the number of 
observed galaxies in it,
which suggests that in order to obtain
a galaxy distribution that
better reproduces
the clustering on large scales, it may be
necessary take into consideration
the environment in which the halo is embedded.
We have further studied two different proxies
for the environment factor -- tidal environment
and smoothed density field, shuffling the
halos within their assigned regions. The result
(shown in Fig. \ref{fig:shuf_rat_cw} 
and Fig. \ref{fig:shuf_rat_den})
is much less discrepant in that case compared
with the full sample, which is another compelling
piece of evidence suggesting that the inclusion
of an environment parameter may be crucial to obtaining
the correct clustering amplitude on large scales.
The correlation between the ages of the galaxies
and the ages of their
dark matter halos has not yet been 
extensively studied in IllustrisTNG
and thus remains an open question. 
If shown to be strong,
it would also have an effect on the observed 
clustering in color-selected samples,
for which there has already been evidence
in the literature. We leave the study
of color-dependent clustering in TNG \citep{TNGbim}
for a subsequent paper, realizing its
potential as a systematic error
that would need to be accounted for in
{redshift space distortions}
constraints coming from future surveys
such as DESI \citep{2013arXiv1308.0847L}. 

Another important factor
that may be at play is the effect of 
baryon physics processes on the matter distribution,
which may contribute significantly to the bias between
the halo and the galaxy distributions.
To illustrate, it is possible that violent processes such as
AGN feedback may expel enough material to cause 
the intrinsic properties of a halo, such as
its concentration, to vary considerably between
the dark-matter only and the hydrodynamical simulations.
If this hypothesis is true, then extracting the properties of
halos from N-body simulations in order to 
generate mock galaxy catalogs might lead to certain issues
and the effect of baryonic physics would need to somehow
be accounted for in the final N-body products.

Since the large-scale clustering obtained from the
IllustrisTNG 300 Mpc simulation box matches
the clustering of real galaxies reasonably well \citep{2018MNRAS.475..676S},
we can conclude from our results
that the basic HOD model for assigning
galaxies to halos in an N-body simulation
introduces too significant of a discrepancy on large scales
to meet the required level of 
accuracy for upcoming experiments. We,
therefore, suggest a possible 
direction for alleviating
this problem -- namely, by 
including secondary halo parameters
(assembly bias parameters), in addition to mass, to the HOD model.
However, the design of a viable model that would improve the precision to the required $\sim 1$\% level
is left for future work. 

The availability of an even larger 
hydrodynamical galaxy formation simulation would be 
extremely beneficial to
expanding our knowledge of
the relationship between galaxies
and their dark matter halos.
It would not only allow us to
check and verify the results 
obtained with TNG300-1,
but also provide us with
substantially more
objects. We would 
have a sufficiently large
dataset to draw conclusions about 
the large-scale structure of the 
Universe with a high degree of 
confidence, possibly including 
tertiary parameters to
the HOD model to capture the 
behavior even better.
Finally, we hope that it would
open the doors for creating 
improved
HOD models that recover
the galaxy clustering on large 
scales with subpercentage
precision -- a feat that could bridge important
gaps in light of future galaxy surveys. 

\section*{Acknowledgements}

We thank Volker Springel, {Benedikt Diemer, Kai Hoffman  and the referee 
Darren Croton} for their illuminating and useful
comments. The IllustrisTNG data used in this 
paper is stored on the FASRC Cannon cluster supported by the FAS 
Division of Science Research Computing Group at Harvard University.
DJE is supported by U.S. Department of Energy 
grant DE-SC0013718 and as a Simons Foundation Investigator.
The Flatiron Institute is supported by the Simons Foundation.




\bibliographystyle{mnras}
\bibliography{refs} 





\bsp	
\label{lastpage}
\end{document}